\documentclass[aip,jcp,reprint]{revtex4-2}  %
\usepackage[utf8]{inputenc}
\usepackage[
  colorlinks=true,
  linkcolor=red,    %
  citecolor=blue,   %
  urlcolor=blue    %
]{hyperref}
\usepackage{graphicx, color}
\usepackage{mathtools}
\usepackage{siunitx}
\usepackage{booktabs}
\usepackage{multirow}
\usepackage{amsmath}
\usepackage{braket}
\usepackage{mycommands}
\usepackage{float}
\usepackage{mhchem}
\usepackage{amsfonts}
\usepackage{array}
\usepackage{CJKutf8} %
\usepackage{soul}

\newcommand{\Refx}[1]{Ref.~\onlinecite{#1}}
\renewcommand{\vec}[1]{\mathbf{#1}}
\newcommand{\Tab}[1]{Table~\ref{#1}}

\newcommand{\rowtitle}[1]{\multicolumn{1}{l}{#1}}

\begin{document}
\begin{CJK*}{UTF8}{gbsn} %

\title{Exact tunneling splittings of rotationally excited states from symmetrized path-integral molecular dynamics}

\author{L\'{e}a Zupan}
\altaffiliation{On exchange from \'Ecole Polytechnique F\'ed\'erale de Lausanne (EPFL)} %
\author{Yu-Chen Wang (汪宇晨)}
\email{wangyuc@phys.chem.ethz.ch}
\author{Jeremy O. Richardson}
\email{jeremy.richardson@phys.chem.ethz.ch}
\affiliation{\mbox{Institute of Molecular Physical Science, ETH Z\"{u}rich, 8093 Z\"{u}rich, Switzerland}}

\date{\today}

\begin{abstract}
We extend our previous symmetrized path-integral molecular dynamics approach to calculate tunneling splittings of molecules in rotationally excited states.
In this new formalism, the system is rigorously projected onto selected rotational manifolds and
states of a chosen symmetry
through an Eckart spring, which connects the two end beads of the ring polymer via %
a permutation--inversion--rotation operation.
This method is numerically exact within statistical uncertainty once convergence with respect to all simulation parameters has been achieved.
Importantly, it enables the simultaneous extraction of tunneling splittings for multiple total angular-momentum quantum numbers $J$ from a single set of simulations, %
without additional computational cost relative to the original approach. %
After validating the formalism by computing the rotational levels of water (beyond the rigid-rotor approximation), 
we apply it to ammonia and obtain rotationally resolved tunneling splittings in excellent agreement with exact variational benchmarks.
Except for small errors due to the underlying potential energy surface,
the results capture the experimentally observed trend that the tunneling splitting decreases with $J$.
\end{abstract}

\maketitle
\end{CJK*}

\section{Introduction}

Quantum tunneling is a fundamental phenomenon in molecular systems that enables atoms to traverse potential energy barriers that are classically forbidden.\cite{Hund1927tunnel}
It plays a significant role in many molecular processes, for instance by increasing the rates of both adiabatic and nonadiabatic chemical reactions.\cite{BellBook,SCHREINER2020980,GRperspective} %
Additionally, tunneling between equivalent molecular configurations %
lifts the degeneracy of symmetry-related states, resulting in energy-level splittings. These splittings can be measured using various spectroscopic techniques, including high-resolution microwave\cite{GordyCookBook,Urban1984_NH3,hexamerprism} and infrared\cite{Liu1996clusters,Birer2009review} spectroscopy, as well as magnetic resonance,\cite{Simenas2020tunnel}
some of which are capable of resolving the splitting at the level of individual rotational states.
Owing to their extreme sensitivity to the height and shape of the barrier, tunneling splittings serve as powerful probes of the underlying potential energy surface (PES), providing information not just about the minima but also %
about the barrier regions involved in the tunneling process. \cite{%
Liu1996clusters,StoneBook}

A variety of theoretical approaches have been developed to calculate tunneling splittings in molecular systems.
Wavefunction-based methods \cite{Schroeder2011malonaldehyde,Hammer2011malonaldehyde,Carrington2017quantum,Lauvergnat2023malonaldehyde,Simko2025trimer} obtain energy levels by solving the Schr\"{o}dinger equation variationally\footnote{We use the term `variational' in the loose sense employed in \Refx{Carrington2017quantum}} in a finite basis. While these methods can in principle provide highly accurate results, their applicability is limited by the steep computational cost that grows rapidly with system size.
Additionally, treating rotationally excited states is substantially more complicated (both formally and numerically) than the rotational ground state ($J=0$), as rovibrational coupling terms\cite{Watson_RotVibHamilt} no longer vanish for $J>0$ and must be treated explicitly.

To circumvent the curse of dimensionality, the diffusion Monte Carlo (DMC) method can be applied, which %
uses random walkers to sample the ground- and excited-state wavefunctions.\cite{Suhm1991DMC,Quack1995DMC,Gregory1995dmc,Mizukami2014malonaldehyde}
However, as DMC requires separate calculations for these two states in order to determine the tunneling splitting, achieving convergence can be difficult when their energy difference is small. %
Furthermore, practical DMC treatments of excited states typically %
require the construction of state-specific nodal surfaces. %
This inevitably introduces systematic errors that are difficult to estimate and makes calculations of  rotationally excited tunneling splittings particularly challenging,\cite{petit_diffusion_2009} especially in highly fluxional systems where rotation--vibration separation breaks down.\cite{hinkle_diffusion_2011,schmiedt_symmetry_2015,wodraszka_ch5_2015}

A more efficient (albeit approximate) method for computing tunneling splittings is provided by instanton theory,\cite{Uses_of_Instantons,Benderskii,Milnikov2008review,tunnel,Perspective,InstReview,Erakovic2020instanton}
which is derived from the semiclassical limit of the path-integral formalism\cite{Miller1971density}
and can be systematically improved by including perturbative corrections.\cite{AnharmInst}
Although instanton theory has been used successfully in many molecules and clusters,\cite{water,hexamerprism,formic,Cvitas2018instanton,i-wat2,pentamer,asymtunnel,chiral,tropolone,asymtunnel,Erakovic2021trimer,Tokic2025hexamer}
achieving quantitative accuracy for weakly-bound dimers \cite{TransferLearning2} or for tunneling through low barriers remains challenging.\cite{Videla_protontrans}
Suggestions have also been proposed for treating rotationally excited tunneling splittings within instanton theory,\cite{Vaillant2018instanton}
but quantitative accuracy has not yet been achieved, 
and such approaches could benefit from
further theoretical development and more extensive benchmarking.

Path-integral molecular dynamics (PIMD) offers a promising route to overcome the limitations discussed above.
In this approach, the quantum system is mapped onto a classical isomorphic ring-polymer representation, enabling efficient sampling using classical techniques and favorable scaling to larger systems.\cite{Ceperley1987exchange,Alexandrou1988tunnelling,Marchi1991tunnelling}
In principle, numerically exact energy differences between the lowest-lying states within specified symmetry subspaces can be obtained using PIMD.\cite{Matyus2016tunnel2}
In contrast to DMC, where tunneling splittings are obtained from separately computed energies, PIMD directly evaluates the splitting. %
As a result, both large and small  splittings can be determined with comparable relative statistical uncertainties.
However, earlier versions of PIMD were applied to calculate tunneling splittings in molecular systems without explicit %
projection onto the rotational ground state.\cite{Matyus2016tunnel1,Vaillant2018dimer,Vaillant2019water,Zhu2022trimer}
As a result, the tunneling splittings obtained can be significantly contaminated by contributions from rotational excitation.

Recently, 
by introducing symmetrized partition functions and 
averaging over orientations via an ``Eckart spring,"
we reformulated PIMD to rigorously isolate the $J=0$ manifold and thereby enabled the calculation of the exact ground-state tunneling splitting.\cite{PIMDtunnel}
This approach was later applied to malonaldehyde and yielded a highly accurate result with an error bar lower than any previous variational or DMC calculations.\cite{malonaldehydePIMD}
It has also been recently applied to \ce{(HCl)2} using a new ab initio PES.\cite{Shen-JCP-2025-154307}
Nevertheless, an extension to rotationally excited states remains to be developed.

In this work, the symmetrized PIMD approach is extended to allow projection onto states with a specified total angular-momentum quantum number $J$.
While retaining the same underlying numerical implementation including the Eckart spring, this extension enables the calculation of exact tunneling splittings for rotationally excited states with the same computational effort as for $J=0$.  In fact, it allows a single set of PIMD simulations to provide results for all $J$ values of interest through post-processing---a capability not previously possible with traditional methods.

The remainder of this paper is structured as follows. 
Section~\ref{sec:Theory} develops the central theoretical framework for obtaining tunneling splittings from symmetrized partition functions, treating the ground and excited rotational states on an equal footing.
Section~\ref{sec:methods} describes the PIMD implementation for computing rotationally-projected partition-function ratios and the resulting tunneling splittings.
Section~\ref{sec:Results} benchmarks the new method on water and ammonia against wavefunction-based variational calculations and experimental measurements.
Section~\ref{sec:conclusion} provides further discussion and concluding remarks.

\section{Theory}\label{sec:Theory}
Within the Born–Oppenheimer approximation and in the absence of external fields, we consider a closed-shell molecular system, neglecting hyperfine interactions.
Under these assumptions, the system is described by a rovibrational Hamiltonian $\hat{H}$ that is invariant under spatial rotations.
Our goal is to use path-integral simulations to extract energy differences between the lowest states belonging to selected rotational and symmetry subspaces.
Following the approach introduced in \Refx{PIMDtunnel}, this can be achieved by applying symmetry-adapted projections to the canonical partition function %
to isolate the contributions of individual quantum states in the low-temperature limit. %
In the following, we first address the projection onto a specified rotational manifold and then extend the methodology to incorporate the discrete permutation--inversion symmetry of the molecule.

\subsection{Rotational projection} \label{subsec:proj_rot_states}
The rotational invariance of the molecular Hamiltonian implies that its eigenstates can be classified according to the total angular-momentum quantum number $J$ and its space-fixed projection $M=-J,\dots,J$.\cite{Bunker,BunkerBook}
We therefore use the set of quantum numbers $(n,J,M)$ to label any eigenstate of the system, where $n$ collectively denotes the remaining rovibrational structure (including vibrational and tunneling character) that cannot be resolved into separate quantum numbers in the presence of rovibrational coupling.
In this representation, the canonical partition function can be written as
\begin{subequations}
\begin{align}
    Z &= \Tr[\mathrm{e}^{-\beta\hat{H}}] \\
        &=\sum_{n,J,M} \braket{n,J,M | \mathrm{e}^{-\beta \hat{H}} | n,J,M}
    \\ &= \sum_{n,J} (2J+1) \, \eu{-\beta E_{n,J}} ,
\end{align}
\end{subequations}
where $\beta = 1/k_\mathrm{B}T$ is the inverse temperature and $E_{n,J}$ denotes the energy level associated with $n$ and $J$.
Note that we have accounted for the degeneracy of the states that differ only in $M$. %

To isolate the contribution from the manifold of states with total angular momentum $J$, we define the $J$-projected partition function by averaging over the degenerate $M$ components:
\begin{subequations}
\label{eq:zj-def}
\begin{align}
     Z^{(J)} &= \frac{1}{2J+1} \sum_{n,M} \braket{n,J,M|\mathrm{e}^{-\beta \hat{H}}|n, J, M}\\
     &= \sum_{n}\mathrm{e}^{-\beta E_{n,J}}.
\end{align}
\end{subequations}
This contains only the spectral information associated with the $J$ manifold, %
but it is not useful in practice as it requires explicit knowledge of the energy levels.
To make progress, we introduce the rotated partition function
\begin{equation}
       Z(\boldsymbol{\Omega}) = \Tr[\mathrm{e}^{-\beta \hat{H}}\hat{R}(\boldsymbol{\Omega})] ,
       \label{eq:zomega}
\end{equation}
where $\hat{R}(\boldsymbol{\Omega})$ represents an overall rotation about the center of mass and is parameterized by $\boldsymbol{\Omega}$ with respect to a space-fixed frame.\footnote{ $\boldsymbol{\Omega}$ can represent a set of Euler angles or any other suitable parametrization of molecular orientation}
Expanding the trace in the eigenbasis gives
\begin{subequations} \label{eq:Rot_partfunc}
\begin{align}
    Z(\boldsymbol{\Omega}) &= \sum_{n,J,M} \braket{n,J,M | \mathrm{e}^{-\beta \hat{H}}\hat{R}(\boldsymbol{\Omega}) | n,J,M} \\
    &= \sum_{n,J,M} \eu{-\beta E_{n,J}}  \braket{n,J,M | \hat{R}(\boldsymbol{\Omega}) | n,J,M} \\
    &= \sum_{n,J,M} D^{(J)}_{MM}(\boldsymbol{\Omega}) \, \eu{-\beta E_{n,J}} ,
\end{align}
\end{subequations}
where $D^{(J)}_{MM}(\boldsymbol{\Omega}) $ is a diagonal element of a Wigner $D$-matrix corresponding to the total angular momentum $J$ and its projection $M$. \cite{ZareBook}

The Wigner $D$-matrices obey the following orthogonality relation:\cite{ZareBook} %
\begin{equation}\label{eq:orthogonality}
    \int \mathrm{d} \boldsymbol{\Omega} \, D_{M'M'}^{(J')}(\boldsymbol{\Omega})^* \,D_{MM}^{(J)}(\boldsymbol{\Omega}) = \frac{8 \pi^2}{2J + 1} \delta_{J'J}\delta_{M'M} .
\end{equation}
This enables the contribution of a specific rotational manifold to the canonical partition function to be isolated.
In particular, averaging%
\footnote{In principle, the orthogonality relation would allow us to pick out states with particular values of $M$.
From symmetry arguments, however, we know that states differ only by their value of $M$ are degenerate. %
As such, we prefer to define a rotationally projected partition function that averages over the degenerate $M$ states, as we found that the average leads to a more stable numerical convergence of the resulting PIMD method. %
}
over $M$ leads to the compact expression\footnote{Since the trace of the Wigner $D$-matrix is real-valued for all $\boldsymbol{\Omega}$, the complex conjugate in Eq.~\eqref{eq:14cZ_proj_rot} is omitted.}
\begin{equation}
  Z^{(J)} =\frac{1}{8\pi^2} \int \mathrm{d} \boldsymbol{\Omega} \, \mathstrut \Tr[D^{(J)}(\boldsymbol{\Omega})] \, Z(\boldsymbol{\Omega}), \label{eq:14cZ_proj_rot} 
\end{equation}
where the trace of the Wigner $D$-matrix is given by %
\cite{varshalovich1988quantum}
\begin{equation} \label{eq:TrD}
     \Tr[D^{(J)}(\boldsymbol{\Omega})]  =  \frac{\sin\left((2J+1)\frac{\theta}{2}\right)}{\sin\left(\frac{\theta}{2}\right)} ,
\end{equation}
with the effective angle $\theta\in[0,\pi]$ determined directly from the trace of the rotation matrix:
\begin{equation}
    \cos (\theta) = \frac{1}{2}\left( \Tr[\hat{R}(\boldsymbol{\Omega})] - 1\right).
\end{equation}

Equation~\eqref{eq:14cZ_proj_rot} establishes an %
alternative definition of the $J$-projected partition function that does not require explicit knowledge of the energy levels.
As will be shown in Sec.~\ref{sec:methods}, this relation provides a practical route for evaluating $Z^{(J)}$ within the PIMD framework.
It is also instructive to note that the Wigner $D$-matrix reduces to unity for the case of $J=0$, such that Eq.~\eqref{eq:14cZ_proj_rot} exactly recovers the form derived in \Refx{PIMDtunnel}.

\subsection{Symmetrization}
In addition to rotational invariance, the molecular Hamiltonian is also invariant under a discrete set of permutation--inversion operations that form the molecular symmetry group.\cite{Bunker,BunkerBook}
The associated symmetry labels provide an additional classification of the eigenstates beyond the total angular-momentum quantum numbers.
To incorporate these discrete symmetries into the present framework, one can insert symmetry operations into the partition function in a manner analogous to the construction above.
This strategy was proposed in \Refx{PIMDtunnel} to isolate eigenstates of different symmetry for $J=0$.
Here, we combine the same treatment with the rotational projection derived above.

Following \Refx{PIMDtunnel}, for a specific symmetry operation $\hat{P}$, we define the symmetrized partition function %
\begin{subequations} \label{eq:zp-chi}
\begin{align}
    Z_P &= \Tr[\mathrm{e}^{-\beta\hat{H}}\hat{P}]
    \\ &= \sideset{}{'}\sum_{n}\sum_{J} (2J+1)\chi^{(n,J)}_P \, \eu{-\beta E_{n,J}} .
\end{align}
\end{subequations}
Here, $\chi_P^{(n,J)}$ is the character of $\hat{P}$ in the irreducible representation (irrep) to which the energy level $(n,J)$ belongs.
The prime on the first summation sign indicates that the sum over $n$ includes only one representative from each degenerate multiplet arising from the discrete permutation--inversion symmetry.
This expression follows by taking the trace in an eigenbasis of $\hat{H}$, in which $\hat{P}$ is block diagonal, with the trace of each block equal to the corresponding character.  As before, the factor $2J+1$ accounts for the degeneracy in $M$.

The quantities $Z_P$ thus encode the symmetry information carried by the eigenstates, and by taking linear combinations, they can be used to project onto a set of states with a chosen discrete symmetry.
However, $Z_P$ still contains contributions from all rotational states.
To isolate the component from a particular rotational manifold, we introduce the $J$-projected symmetrized partition function
\begin{subequations}
\label{eq:zpj-def}
\begin{align}
     Z_{P}^{(J)} &= \frac{1}{2J+1} \sum_{n,M} \braket{n,J,M|\mathrm{e}^{-\beta \hat{H}}\hat{P}|n, J, M} \\
     &=  \sideset{}{'}\sum_{n} \chi^{(n,J)}_P  \, \eu{-\beta E_{n,J}} .
\end{align}
\end{subequations}
This quantity plays the same role here as $Z_P$, %
but it is now resolved with respect to the quantum number $J$.
In the special case where $\hat{P}$ is the identity operation, $Z_P^{(J)}$ reduces to the rotationally projected partition function $Z^{(J)}$ introduced in Eq.~\eqref{eq:zj-def}.

To connect $Z_P^{(J)}$ with quantities accessible in PIMD simulations, we proceed in direct analogy with Sec.~\ref{subsec:proj_rot_states} and introduce a rotated and symmetrized partition function
\begin{equation}
    Z_P(\boldsymbol{\Omega}) = \Tr[\mathrm{e}^{-\beta\hat{H}}\hat{P}\hat{R}(\boldsymbol{\Omega})] .
    \label{eq:zp-rot}
\end{equation}
Using the same orthogonality relations [Eq.~\eqref{eq:orthogonality}] as before, we arrive at
\begin{equation}
    Z_P^{(J)} = \frac{1}{8\pi^2}\int \mathrm{d}\boldsymbol{\Omega} \, \Tr[D^{(J)}(\boldsymbol{\Omega})] \, Z_P(\boldsymbol{\Omega}) ,\label{eq:projectedpartfunc}
\end{equation}
which can be regarded as an extension of Eq.~\eqref{eq:14cZ_proj_rot} to include a projection onto an arbitrary symmetry operation.
Again, for $J=0$, this exactly recovers the form obtained in \Refx{PIMDtunnel}.

These partition functions can be directly related to the energy-level splittings of interest.
In particular, consider the states $(n,J)$ and $(n',J')$, each of which is the lowest-lying state of its irrep within the specified rotational manifold.
Following the same arguments as in \Refx{PIMDtunnel}, and taking the low-temperature limit in which vibrational excitations are exponentially suppressed, the energy difference between the two states can be written as
\begin{align}
    E_{n,J} - E_{n',J'}
    = \frac{1}{\beta}\ln\left[\frac{\sum_{\alpha}g_{\alpha}\chi_{\alpha}^{(n',J')*}Z^{(J')}_{\alpha}}{\sum_{\alpha}g_{\alpha}\chi_{\alpha}^{(n,J)*}Z^{(J)}_{\alpha}} \right] ,
    \label{eq:genTunSplit}
\end{align}
where the sums run over the conjugacy classes $\alpha$ of the molecular symmetry group, $g_\alpha$ is the order of the class, and
$Z^{(J)}_{\alpha}$ is the  $J$-projected symmetrized partition function for a representative symmetry operation $\hat{P}$ in that class.
The calculation of tunneling splittings is therefore reduced to the evaluation of a set of symmetrized partition functions, $Z_P^{(J)}$.

Equations~\eqref{eq:projectedpartfunc} and \eqref{eq:genTunSplit} are the central results of this section.
They constitute a natural extension of the theory in \Refx{PIMDtunnel} to rotationally excited states.
The only difference is that all partition functions involved are now explicitly projected onto a specified rotational manifold rather than restricted to the rotational ground state.
As will be shown in Sec.~\ref{sec:methods}, this allows the PIMD framework developed in \Refx{PIMDtunnel} to be directly inherited by the present formulation without any modifications to the underlying simulation procedure.
Finally, although Eq.~\eqref{eq:genTunSplit} appears to require contributions from each conjugacy class of the symmetry group, in many cases further simplifications are possible, such that only a subset of symmetry operations needs to be considered explicitly, as will be illustrated in Sec.~\ref{sec:Results}.

\section{Methods}\label{sec:methods}
\subsection{Path-integral formalism}
We now express $Z_P^{(J)}$ in the path-integral formalism\cite{Feynman} suitable for PIMD sampling.
Applying the standard PIMD construction\cite{TuckermanBook} to $Z_P(\boldsymbol{\Omega})$  [Eq.~\eqref{eq:zp-rot}] and substituting the result into Eq.~(\ref{eq:projectedpartfunc}), one obtains
\begin{align}
    Z_P^{(J)} &=  \frac{1}{(2\pi \hbar)^{Nf}} \frac{1}{8\pi^2}\int \mathrm{d}\boldsymbol{\Omega} \, \Tr[D^{(J)}(\boldsymbol{\Omega})]\nonumber\\
&\quad\times
  \int \mathrm{d}\boldsymbol{p} \int \mathrm{d}\boldsymbol{r} \,
    \exp[-\beta_N H_{P}(\boldsymbol{p}, \boldsymbol{r};\boldsymbol{\Omega})].
\end{align}
Here, $\boldsymbol{r}=\{\vec{r}^{(i)}\}_{i=1}^N$ and $\boldsymbol{p}=\{\vec{p}^{(i)}\}_{i=1}^N$ denote the Cartesian coordinates and conjugate momenta of the $N$ ring-polymer beads, where  $\vec{r}^{(i)}=\{\vec{r}_a^{(i)}\}_{a=1}^{n_\mathrm{at}}$ and $\vec{p}^{(i)}=\{\vec{p}_a^{(i)}\}_{a=1}^{n_\mathrm{at}}$ are the Cartesian coordinates and momenta of the $i$-th bead.  Additionally, $n_\mathrm{at}$ is the number of atoms,
$f=3n_\mathrm{at}$ is the number of degrees of freedom of the system, and
$\beta_N=\beta/N$ determines the effective inverse temperature of each bead.
In addition to the usual dependence on $\boldsymbol{r}$ and $\boldsymbol{p}$, the ring-polymer Hamiltonian also depends on the rotation, $\boldsymbol{\Omega}$, and the symmetry operator, $\hat{P}$, as
\begin{equation} \label{eq:HNPOmega}
    H_{P}(\boldsymbol{p}, \boldsymbol{r};\boldsymbol{\Omega}) = 
     H_{N}^{(\text{open})}(\boldsymbol{p}, \boldsymbol{r}) + 
     U_{P}(\boldsymbol{r};\boldsymbol{\Omega}),
\end{equation}
with
\begin{align}
    H_{N}^{(\mathrm{open})}(\boldsymbol{p}, \boldsymbol{r}) &= 
        \sum_{i=1}^N \left[ \sum_{a=1}^{n_{\text{at}}} \frac{||\vec{p}_a^{(i)}||^2}{2m_a} + V(\vec{r}^{(i)})\right] \nonumber\\ &\quad+ \sum_{i=1}^{N-1}  \sum_{a=1}^{n_{\text{at}}} \frac{m_a \omega_N^2}{2} \left\lVert \vec{r}_a^{(i+1)} - \vec{r}_a^{(i)} \right\rVert^2
\end{align}
and
\begin{equation}
    U_{P}(\boldsymbol{r};\boldsymbol{\Omega})
    =\sum^{n_\mathrm{at}}_{a=1}\frac{m_a \omega_N^2}{2} \left\lVert \left[\hat{P}^{-1}\vec{r}^{(N)} - \hat{R}(\boldsymbol{\Omega})\vec{r}^{(1)}\right]_a \right\rVert^2 ,
\end{equation}
where $V(\vec{r}^{(i)})$ is the physical potential of the $i$-th bead, $\omega_N= 1/\beta_N \hbar$ is the frequency of the harmonic springs connecting adjacent beads, and $m_a$ is the mass of the $a$-th atom.

In Eq.~\eqref{eq:HNPOmega}, the first term $H_{N}^{(\mathrm{open})}(\boldsymbol{p}, \boldsymbol{r}) $ is the standard Hamiltonian of an open-chain polymer arising from the Trotter discretization of the imaginary-time propagator.\cite{TuckermanBook}
It is  independent of the rotation, $\boldsymbol{\Omega}$, and is identical to the ring-polymer Hamiltonian appearing in traditional PIMD, except that the harmonic spring connecting the two end beads is absent.

The second term $U_{P}(\boldsymbol{r};\boldsymbol{\Omega})$ plays the role of a boundary contribution. %
It replaces the usual closing spring of the ring polymer by connecting bead $N$, after the action of $\hat{P}^{-1}$, to a rotated configuration of bead 1.
When the rotation is trivial and $\hat{P}$ corresponds to the identity operation,  this boundary term reduces to the standard spring connecting the two end beads, and Eq.~\eqref{eq:HNPOmega} recovers the standard ring-polymer Hamiltonian in traditional PIMD.

The remaining complication is the explicit integration over $\boldsymbol{\Omega}$, which originates from the rotational projection.
While the phase-space integral over $(\boldsymbol{p}, \boldsymbol{r})$ can be sampled by standard molecular dynamics, evaluating the additional rotational integral for every polymer configuration along the trajectory would be computationally impractical.
It is important to note, however, that the rotation enters only through the boundary term via the mass-weighted squared distance between $\hat{P}^{-1}\vec{r}^{(N)}$ and the rotated configuration $\hat{R}(\boldsymbol{\Omega})\vec{r}^{(1)}$.
For a given pair of end-bead geometries, this distance is minimized by the optimal rotation $\widetilde{\boldsymbol{\Omega}}$ that satisfies the Eckart conditions.\cite{EckartCondition}

As the number of beads increases, the spring constant of the boundary term becomes progressively stiffer.
Consequently, the weight $\exp[-\beta_NU_{P}]$ for a given ring-polymer configuration is sharply peaked around $\widetilde{\boldsymbol{\Omega}}$, and contributions from other orientations are exponentially suppressed.
The integral over $\boldsymbol{\Omega}$ is therefore dominated by a small neighborhood around $\widetilde{\boldsymbol{\Omega}}$ and can be evaluated by a steepest-descent treatment,\cite{BenderBook} which becomes exact in the limit $N\rightarrow\infty$.
This strategy, originally employed in \Refx{PIMDtunnel} for the $J=0$ case, can be %
easily generalized for any $J$, leading to

\begin{equation}
    Z_P^{(J)} = (2 \pi \hbar)^{-Nf}\int \mathrm{d}\boldsymbol{p}\int \mathrm{d}\boldsymbol{r} \, u_{P}^{(J)}(\boldsymbol{r}) \, \mathrm{e}^{-\beta_N\widetilde{H}_{P}(\boldsymbol{p},\boldsymbol{r})},\label{eq:finalPIMD_Z}
\end{equation}
with the effective Hamiltonian
\begin{equation}
\label{eq:hnp-final}
    \widetilde{H}_{P}(\boldsymbol{p}, \boldsymbol{r}) = 
     H_{N}^{(\text{open})}(\boldsymbol{p}, \boldsymbol{r}) + 
     U_{P}(\boldsymbol{r};\widetilde{\boldsymbol{\Omega}})
\end{equation}
and a prefactor
\begin{align}
    u^{(J)}_{P}(\boldsymbol{r}) &= \frac{1}{8\pi^2} \left( \frac{2\pi}{\beta_N\omega_N^2}\right)^{3/2}
    \Tr[D^{(J)}(\widetilde{\boldsymbol{\Omega}})] \nonumber\\
    &\quad\times\left(\det\boldsymbol{\Theta}[\hat{R}(\widetilde{\boldsymbol{\Omega}})\vec{r}^{(1)},\hat{P}^{-1}\vec{r}^{(N)}]\right)^{-1/2} .
    \label{eq:Eckpref}
\end{align}
Here, the second line of Eq.~\eqref{eq:Eckpref} arises from the fluctuation contribution associated with the steepest-descent evaluation of the integral over $\boldsymbol{\Omega}$.
The $3 \times 3$ matrix $\boldsymbol{\Theta}[\vec{r},\vec{r}']$ is a function of the two bead configurations $\vec{r}$ and $\vec{r}'$, with elements given by
\begin{equation}
    \boldsymbol{\Theta}_{\mu\nu}[\vec{r},\vec{r}'] = \sum_{a=1}^{n_\mathrm{at}} m_a \left[ (\vec{r}_a \cdot \vec{r}_a')\delta_{\mu\nu} - r_{a,\mu}r_{a,\nu}'\right],
\end{equation}
where $\mu,\nu\in(x,y,z)$.

The boundary term $ U_{P}(\boldsymbol{r};\widetilde{\boldsymbol{\Omega}})$ in Eq.~\eqref{eq:hnp-final} is the Eckart spring introduced in \Refx{PIMDtunnel}.
Note that both $ U_{P}(\boldsymbol{r};\widetilde{\boldsymbol{\Omega}})$ and the trace $\Tr[D^{(J)}(\widetilde{\boldsymbol{\Omega}})]$ in Eqs.~\eqref{eq:hnp-final} and \eqref{eq:Eckpref} are now evaluated at the optimal rotation $\widetilde{\boldsymbol{\Omega}}$, which is itself an implicit function of the end-bead configurations.
In practice, this optimal rotation can be obtained using an efficient algorithm based on quaternion algebra.\cite{Krasnoshchekov2014Eckart}
The final form Eq.~\eqref{eq:finalPIMD_Z} is therefore expressed entirely in terms of the phase-space variables $(\boldsymbol{p}, \boldsymbol{r})$, making it directly amenable to standard PIMD sampling.
Appendix~\ref{app:EckartForce} provides a detailed derivation of the forces associated with the Eckart spring for permutation, inversion, and permutation--inversion symmetry operations, as required for the PIMD propagation.

It is worth emphasizing that the effective Hamiltonian, $\widetilde{H}_{P}$, obtained here is identical to that derived in \Refx{PIMDtunnel} for the $J=0$ case.
In particular, it is independent of the total angular-momentum quantum number $J$.
The only modification introduced by the present extension is the appearance of the factor $\Tr[D^{(J)}(\widetilde{\boldsymbol{\Omega}})]$ in the prefactor $u_P^{(J)}(\boldsymbol{r})$.

As a consequence, the PIMD implementation developed in \Refx{PIMDtunnel} (and related algorithms)\cite{Yuchen-PIHMC-2026} can be directly employed here to propagate the trajectories.
For a given symmetry operation $\hat{P}$, a single PIMD simulation therefore provides all the information required to evaluate $Z_P^{(J)}$ for different values of $J$, with the dependence on $J$ arising exclusively from the associated prefactors, which can all be computed from the same trajectory.

\subsection{Ensemble averages and free-energy differences}

While Eq.~(\ref{eq:finalPIMD_Z}) provides an explicit expression for $Z_P^{(J)}$, a direct evaluation via PIMD is usually impractical, because PIMD evaluates thermal averages under a given distribution rather than partition functions themselves.
This issue can be addressed by rewriting the energy-splitting expression in Eq.~(\ref{eq:genTunSplit}) in a form that involves only ratios of partition functions, which is easily achieved by dividing both numerator and denominator by a suitable reference partition function.
These ratios
can be recast into the form
\begin{equation}
\label{eq:zpj-ratio}
    \frac{Z_{P'}^{(J')}}{Z_{P}^{(J)}} = \frac{\langle u_{P'}^{(J')}(\boldsymbol{r})\rangle_{P'}}{\langle u_{P}^{(J)}(\boldsymbol{r})\rangle_{P}} \, \mathrm{e}^{-\beta_N \Delta F} ,
\end{equation}
where $\langle\cdot\rangle_{P}$ denotes the thermal average with respect to $\widetilde{H}_{P}$, and $\Delta F$
is the free-energy difference between the two effective Hamiltonians, defined via
\begin{equation}
    \mathrm{e}^{- \beta_N \Delta F} = \frac{\int \mathrm{d}\boldsymbol{p} \int \mathrm{d} \boldsymbol{r} \,\exp[-\beta_N\widetilde{H}_{P'}(\boldsymbol{p},\boldsymbol{r})]}{\int \mathrm{d} \boldsymbol{p} \int \mathrm{d} \boldsymbol{r} \,\exp[-\beta_N\widetilde{H}_{P}(\boldsymbol{p},\boldsymbol{r})]} .
    \label{eq:freeE2ratio}
\end{equation}

While the thermal averages are available directly from PIMD simulations, the calculation of the free-energy difference, $\Delta F$, is more challenging.
Nonetheless, a variety of approaches developed for classical statistical mechanics can be used here.\cite{TuckermanBook,MolSim}
In this work, we evaluate $\Delta F$ by thermodynamic integration\cite{TI_originalpaper,MolSim} (TI) along a continuous thermodynamic path defined by a parameter $\xi \in [0,1]$, which smoothly connects the two end-points.
In particular, we introduce a family of effective Hamiltonians, %
\begin{equation}
\widetilde{H}_{\xi} = (1 - \xi)\widetilde{H}_{P} +\xi \widetilde{H}_{P'},
\label{eq:Hamiltonian_TI}
\end{equation}
with partition functions
\begin{equation}
Z_{\xi} = (2 \pi \hbar)^{-Nf} \int \mathrm{d} \boldsymbol{p} \int \mathrm{d} \boldsymbol{r}\, \mathrm{e}^{-\beta_N \widetilde{H}_{\xi}(\boldsymbol{p},\boldsymbol{r})}.
\end{equation}
The corresponding free-energy difference is then rewritten as
\begin{subequations}
\begin{align}
    \Delta F &= - \frac{1}{\beta_N} \int_0^1 \frac{\mathrm{d} \ln Z_{\xi}}{\mathrm{d} \xi}\,\mathrm{d} \xi \\
    &=\int_0^1 \left\langle \widetilde{H}_{P'} - \widetilde{H}_{P}\right\rangle_{\xi}\, \mathrm{d}\xi ,
    \label{eq:DeltaF_TIintegral}
\end{align}
\end{subequations}
where $\langle\cdot\rangle_\xi$ represents the thermal average with respect to the intermediate Hamiltonian $\widetilde{H}_\xi$.

In practice, the integral over $\xi$ in Eq.~(\ref{eq:DeltaF_TIintegral}) is evaluated using the shifted Gauss--Legendre (sGL) quadrature scheme.\cite{NumRep}
In particular, the thermal averages in Eq.~(\ref{eq:DeltaF_TIintegral}) are computed with PIMD simulations at each quadrature point $\xi_i$, and then these averages are combined with the corresponding sGL weights to estimate the integral.
Note that the PIMD simulations at the TI end-points ($\xi=0$ and $\xi = 1$) can additionally be used to calculate the thermal averages of the Eckart-spring prefactors $\braket{u_{P}^{(J)}}_{P}$ and $\braket{u_{P'}^{(J')}}_{P'}$, which are required to obtain the ratios in Eq.~\eqref{eq:zpj-ratio}. %
Finally, it is worth highlighting that $\Delta F$ is independent of $J$, so that a single TI calculation can be reused to obtain partition-function ratios for multiple values of $J$ simultaneously.

\section{Results and Discussion}\label{sec:Results}
We now apply the newly developed approach to compute the rotational energy levels of the water molecule and the tunneling splittings of rotationally excited states of ammonia.
To accommodate the stiff ring-polymer spring terms and the more slowly varying physical potential, a multiple-timestep integrator\cite{TuckermanBook} was used. %
In this scheme, the forces associated with the fast-varying ring-polymer spring terms were updated with a smaller timestep $\delta t=\Delta t/8$, while those associated with the slow-varying physical potential were updated with a larger timestep $\Delta t=0.2\,\mathrm{fs}$.
In addition, a thermostat is required to sample the appropriate Boltzmann distribution.
For the systems studied below, an Andersen thermostat\cite{Andersen} was used, 
in which a subset of the degrees of freedom was randomly selected after each step, and the corresponding momenta were resampled independently.
The size of the subset was set such that each momentum component was resampled once every 100\,fs on average.
These parameters were found to be suitable for other molecular systems %
as demonstrated in earlier studies.\cite{malonaldehydePIMD,PIMDtunnel}
Statistical uncertainties in both the intermediate quantities and the final results were estimated following the same bootstrapping procedure as in \Refx{PIMDtunnel}, with further details provided in the Supplementary Material.
Additional computational details specific to each system are given %
below.

\subsection{Water}
We first consider \ce{H2O} as a simple test case.
Although isolated water molecules do not exhibit tunneling splittings, they provide a convenient proof of principle for validating our rotational-projection formulation.

H$_2$O is an asymmetric top with rotational energy levels denoted by $J_{K_aK_c}$, where the conventional labels $K_a$ and $K_c$ correlate with the symmetric-top quantum number in the prolate and oblate limit, respectively.\cite{GordyCookBook}
The molecular symmetry group is isomorphic to $C_\mathrm{2v}$, whose character table is shown in \Tab{tab:C2v}\@.
Here, $E$ is the identity, $(12)$ denotes the permutation of the two hydrogen atoms, $E^*$ denotes an inversion through the center of mass, and $(12)^*$ is the combination of the permutation and inversion operations. %

\begin{table}[htbp]
\caption{Character table for the $C_\mathrm{2v}$ molecular symmetry group of water. Row headings denote conventional irrep labels and column headings correspond to symmetry operations, each of which forms its own conjugacy class.}
\label{tab:C2v}
{\renewcommand{\arraystretch}{1.5}
\begin{ruledtabular}
\begin{tabular*}{\columnwidth}{@{\extracolsep{\fill}}lcccc}
$C_\mathrm{2v}$ & $E$ & $(12)$ & $E^*$ & $(12)^*$ \\
\hline
$A_1$ & $\phantom{-}1$ & $\phantom{-}1$ & $\phantom{-}1$ & $\phantom{-}1$ \\
$A_2$ & $\phantom{-}1$ & $\phantom{-}1$ & $-1$ & $-1$ \\
$B_1$ & $\phantom{-}1$ & $-1$ & $-1$ & $\phantom{-}1$ \\
$B_2$ & $\phantom{-}1$ & $-1$ & $\phantom{-}1$ & $-1$ \\
\end{tabular*}
\end{ruledtabular}
}
\end{table}

\begin{figure}
    \centering
    \includegraphics[width=0.95\linewidth]{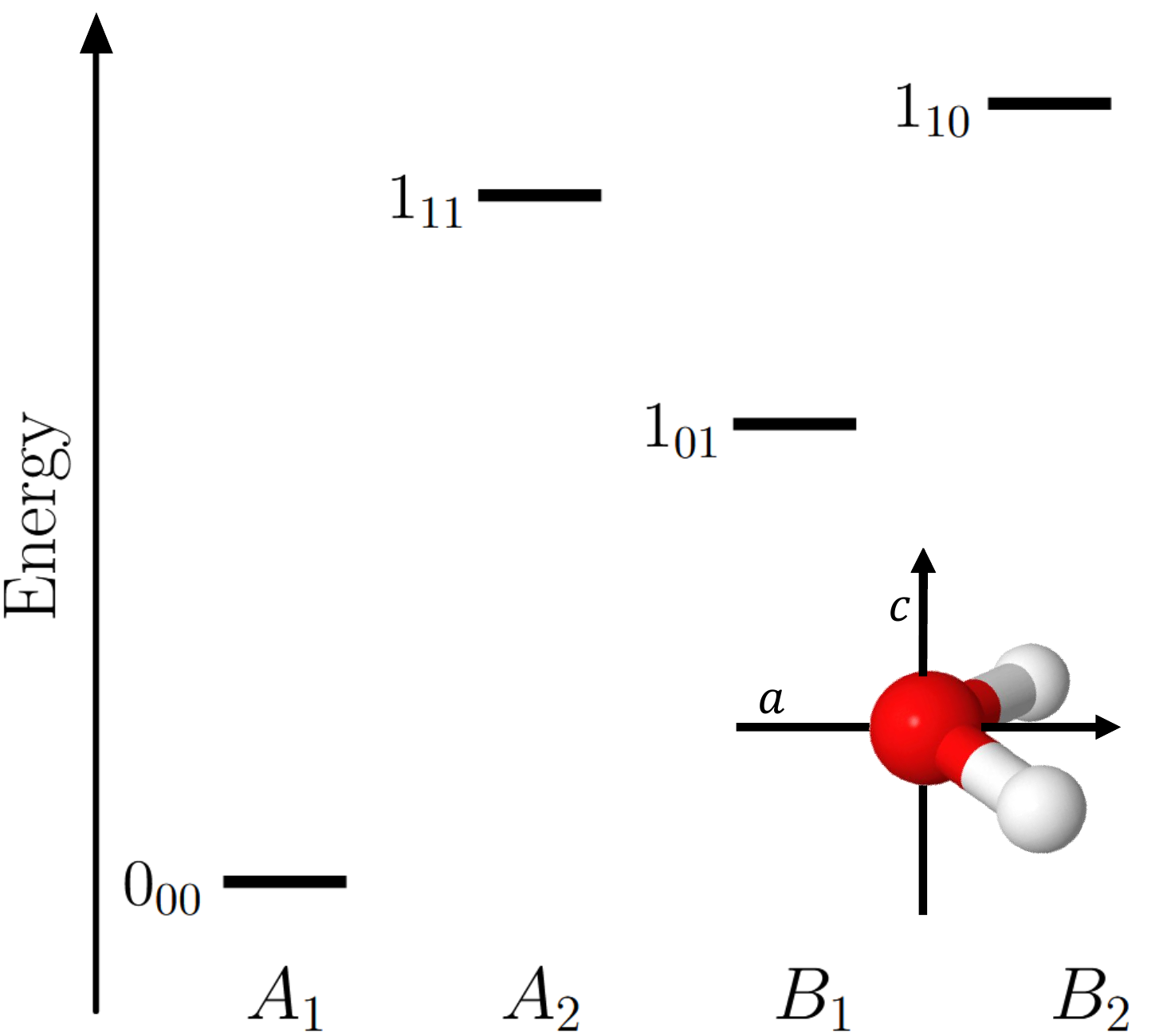}
    \caption{Energy-level diagram of the relevant rotational states of water, labeled by their respective irreps. The states are denoted by $J_{K_aK_c}$ where $K_a$ and $K_c$ are the projections of the angular momentum $J$ onto the $a$ and $c$ principal inertia axes, as indicated in the inset. %
    }
    \label{fig:water}
\end{figure}

The rotational energy levels of  \ce{H2O} in the $J=0$ and $J=1$ manifolds are shown in Fig.~\ref{fig:water}, with each state labeled by its irrep.
Within our framework, the desired energy differences can be obtained from a set of partition-function ratios $Z_{P'}^{(J')}/Z_{P}^{(J)}$, which in turn require both the corresponding Eckart-spring prefactors and free-energy differences.
For \ce{H2O}, however, the latter are not essential.
By exploiting the structure of the low-lying rovibrational spectrum, one can derive a simple relation between $\Delta F$ and the $J=0$ prefactors  in the low-temperature limit, such that the former can be directly obtained from the latter, thereby avoiding expensive TI calculations.

To this end, we focus on the $J=0$ projected partition functions.
As shown in Fig.~\ref{fig:water}, the $J=0$ manifold contains only the $0_{00}$ ground state of $A_1$ symmetry. %
According to Eq.~\eqref{eq:zp-chi} and using \Tab{tab:C2v}, the quantities $Z_P^{(J=0)}$ become identical for all $\hat{P}\in C_\mathrm{2v}$, such that
\begin{equation} \label{ZsJ0}
    Z_E^{(J=0)} = Z_{(12)}^{(J=0)} = Z_{E^*}^{(J=0)} = Z_{(12)^*}^{(J=0)}
\end{equation}
in the low-temperature limit.
We can therefore extract the free-energy difference $\Delta F$ associated with the ratio $Z_{P'}^{(J')}/Z_{P}^{(J)}$ by noting that $Z_{P'}^{(J=0)}/Z_{P}^{(J=0)}$ approaches 1 as $T\to 0$.
In this way, Eq.~\eqref{eq:zpj-ratio} for $J=J'=0$ gives
\begin{equation}
\label{eq:df-from-upj}
\mathrm{e}^{-\beta_N \Delta F} = 
\frac {\langle u_{P}^{(J=0)}(\boldsymbol{r})\rangle_{P}}
 {\langle u_{P'}^{(J=0)}(\boldsymbol{r})\rangle_{P'}}\,.
\end{equation}
Substituting Eq.~\eqref{eq:df-from-upj} into Eq.~\eqref{eq:zpj-ratio}, we obtain
\begin{equation}
\label{eq:zpj-ratio-without-df}
    \frac{Z_{P'}^{(J')}}{Z_{P}^{(J)}} = \frac{\langle u_{P'}^{(J')}(\boldsymbol{r})\rangle_{P'}}{\langle u_{P}^{(J)}(\boldsymbol{r})\rangle_{P}}
\frac {\langle u_{P}^{(J=0)}(\boldsymbol{r})\rangle_{P}}
 {\langle u_{P'}^{(J=0)}(\boldsymbol{r})\rangle_{P'}}\,.
\end{equation}
Using Eq.~\eqref{eq:zpj-ratio-without-df} instead of Eq.~\eqref{eq:zpj-ratio} removes the need for separate free-energy calculations, greatly simplifies the workflow and reduces the computational cost.
Strictly speaking, Eq.~\eqref{eq:zpj-ratio-without-df} is exact only in the low-temperature limit, such that contributions from excited vibrational levels are negligible.
This does not introduce an additional limitation, because our method already relies on the same low-temperature condition to extract energy-level splittings from thermal ensembles.

On this basis, the \ce{H2O} calculations require only four PIMD simulations, one for each symmetry operation $\hat{P}\in\{E,(12),E^*,(12)^*\}$.
Each simulation is performed under the corresponding ring-polymer Hamiltonian $\widetilde{H}_P$ and yields the ensemble averages $\langle u_P^{(J)}(\boldsymbol{r})\rangle_P$ for any $J$ of interest.
The ratios $Z_{P'}^{(J')}/Z_{P}^{(J)}$ are then obtained directly by post-processing these results using Eq.~\eqref{eq:zpj-ratio-without-df}.
We employed the ab initio %
PES developed by Partridge and Schwenke.\cite{PartridgePES}
Each of the simulations consisted of 96 independent trajectories, run in parallel, which were thermalized for 5\,ps, followed by a 25\,ps run for data collection.

We first consider energy differences within the $J=1$ manifold. %
Before presenting the numerical results, it is %
instructive to examine the expressions for the energy-level differences in more detail.
As shown in Fig.~\ref{fig:water}, there are no low-energy states of $A_1$ symmetry in the \mbox{$J=1$} manifold, which implies that in the low-temperature limit where vibrational excitations are frozen out,
\begin{equation}\label{eq:zsum}
    Z_E^{(J=1)} + Z_{(12)}^{(J=1)} + Z_{E^*}^{(J=1)} + Z_{(12)^*}^{(J=1)} = 0 .
\end{equation}
This relationship serves as a useful internal consistency check and allows Eq.~(\ref{eq:genTunSplit}) to be recast into the following compact expressions:
\begin{subequations} \label{eq:H2O_J1}
    \begin{align}
        E_{1_{10}}-E_{1_{01}} &= \frac{1}{\beta} \ln\left(\frac{Z_{E}^{(J=1)}+Z_{(12)^*}^{(J=1)}}{Z_{E}^{(J=1)}+Z_{E^*}^{(J=1)}} \right) \label{eq:H2O_J1_1} ,  \\
    E_{1_{11}}-E_{1_{01}} &= \frac{1}{\beta} \ln\left( \frac{Z_{E}^{(J=1)}+Z_{(12)^*}^{(J=1)}}{Z_{E}^{(J=1)}+Z_{(12)}^{(J=1)}}\right) ,\\
    E_{1_{10}}-E_{1_{11}} &= \frac{1}{\beta} \ln\left(\frac{Z_{E}^{(J=1)}+Z_{(12)}^{(J=1)}}{Z_{E}^{(J=1)}+Z_{E*}^{(J=1)}} \right) .
    \end{align}
\end{subequations}
In order to evaluate these equations in practice, we further divide the numerator and denominator by a reference value, for example $Z_E^{(J=1)}$, so that each expression is written in terms of partition-function ratios.

\begin{table}[htbp]
\caption{Energy differences [in cm$^{-1}$] among the three $J=1$ rotational states of water calculated by PIMD using Eqs.~\eqref{eq:zpj-ratio-without-df} and \eqref{eq:H2O_J1} from simulations performed at 100\,K, with different bead numbers.  Statistical errors are shown in parentheses ($1\sigma$). The boldfaced values are taken as the final PIMD results. For comparison, results from an exact quantum benchmark, experiment and the rigid-rotor approximation are listed in the bottom panel.}
\label{tab:Water-J1-no-dF}
{\renewcommand{\arraystretch}{1.5}
\begin{ruledtabular}
\begin{tabular*}{\columnwidth}{@{\hspace{4pt}}@{\extracolsep{\fill}}cccc}
     $N$  &   {$E_{1_{10}}-E_{1_{11}}$} & {$E_{1_{10}}-E_{1_{01}}$} & {$E_{1_{11}}-E_{1_{01}}$}  \\
\hline
   $32$   &    $5.2(1)$     &   $18.4(2)$   &   $13.2(2)$ \\
   $64$   &    $5.1(1)$     &   $18.4(2)$   &   $13.3(2)$ \\
   $128$  &    $5.2(1)$     &   $18.4(2)$   &   $13.2(2)$ \\
   $256$  &    $\mathbf{5.2(1)}$     &   $\mathbf{18.5(2)}$   &   $\mathbf{13.3(2)}$ \\
\hline
\rowtitle{Variational\cite{PartridgePES}}        &     $5.23$    &       $18.58$        &   $13.34$       \\
\hline
\rowtitle{Experiment\cite{Hall1967H20_rotspec}}    &   $5.224$    &     $18.57(5)$    &   $13.346$   \\
\hline
\rowtitle{Rigid Rotor}     &     $5.06$    &       $17.88$        &   $12.82$       \\
\end{tabular*}
\end{ruledtabular}
}
\end{table}

PIMD simulations were performed at 100\,K, which was chosen to ensure that the partition functions are dominated by the vibrational ground state.
For \ce{H2O}, the lowest vibrational excitation is the bending mode at 1595\,$\text{cm}^{-1}$,\cite{Nielsen_water} so vibrationally excited contributions are expected to be exponentially suppressed at this temperature. 
Note that as we rigorously project onto specified $J$-manifolds, there is no requirement for $k_\mathrm{B}T$ to be smaller than the rotational energy spacing.

The results for the calculated energy differences between the three $J=1$ states are reported in \Tab{tab:Water-J1-no-dF}, together with the corresponding benchmark from a wavefunction-based variational approach, experimental values, and (for comparison) a result based on the rigid-rotor approximation.
The three $J=1$ energy differences obtained from PIMD are statistically consistent for all bead numbers considered, indicating fast convergence with respect to $N$.
We take the $N=256$ values as our final PIMD estimates.
These results agree well with the exact quantum benchmark computed on the same PES, thereby confirming the rationale for the temperature choice discussed above and validating our rotational-projection formulation.
Note that since this PES is empirically adjusted to match the experimental data, both the variational and PIMD results are in excellent agreement with experiment.
By contrast, the rigid-rotor approximation (obtained from the moments of inertia at the equilibrium geometry of \ce{H2O}) underestimates %
the energy differences by 3--4\%.
This emphasizes that the PIMD approach captures the rovibrational coupling that is neglected by the rigid-rotor approximation.

In addition to computing energy differences within a single $J$-manifold, it is also possible to evaluate energy differences between different manifolds.
As before, it is instructive to show that, using Eq.~\eqref{eq:zsum}, the energy differences between the $J=0$ ground state and the $J=1$ rotationally excited states can be obtained as:\footnote{Note that certain symmetry-projected partition functions are negative, such that the expressions inside the logarithms always evaluate to positive numbers.}
\begin{subequations} \label{eq:H2O_J1_0}
\begin{align}
    E_{1_{10}}-E_{0_{00}}  %
    &=\frac{-1}{\beta}\ln\left(-\frac{Z_{(12)}^{J=1}}{2Z_{(12)}^{J=0}} - \frac{Z_{(12)^*}^{J=1}}{2Z_{(12)^*}^{J=0}}\right), \label{eq:H2O_J1_0_c}\\
    E_{1_{01}}-E_{0_{00}} &=\frac{-1}{\beta}\ln\left(-\frac{Z_{(12)}^{J=1}}{2Z_{(12)}^{J=0}} - \frac{Z_{E^*}^{J=1}}{2Z_{E^*}^{J=0}}\right), \\ %
    E_{1_{11}}-E_{0_{00}} &=\frac{-1}{\beta}\ln\left(-\frac{Z_{E^*}^{J=1}}{2Z_{E^*}^{J=0}} - \frac{Z_{(12)^*}^{J=1}}{2Z_{(12)^*}^{J=0}}\right).  %
\end{align}
\end{subequations}
The PIMD results obtained with a range of bead numbers are presented in \Tab{tab:Water_J0_J1_wo_TI}\@.
Again, the converged PIMD results are in good agreement with the quantum-mechanical benchmark.
It is worth noting that the energy-level differences within the $J=1$ manifold reported in \Tab{tab:Water-J1-no-dF} are already converged at $N\ge32$, whereas those between the $J=0$ and $J=1$ manifolds reported in \Tab{tab:Water_J0_J1_wo_TI} require $N\ge128$ for convergence. This indicates that, even for the same system, the convergence with respect to $N$ can differ substantially among different manifolds.
Such behavior is consistent with the well-known observation that convergence with respect to the number of beads is property-dependent.\cite{Berne-ARPC-1986-401}

\begin{table}[htbp]
\caption{The same as \Tab{tab:Water-J1-no-dF}, but for the energy differences between the $J=1$ and $J=0$ manifolds. }
\label{tab:Water_J0_J1_wo_TI}
{\renewcommand{\arraystretch}{1.5}
\begin{ruledtabular}
\begin{tabular*}{\columnwidth}{@{\hspace{4pt}}@{\extracolsep{\fill}}cccc}
   $N$   &   {$E_{1_{10}}-E_{0_{00}}$} & {$E_{1_{11}}-E_{0_{00}}$} & {$E_{1_{01}}-E_{0_{00}}$}  \\
\hline
   $32$    &     $41.6(2)$         &     $36.4(2)$      &    $23.2(1)$  \\
   $64$    &     $41.8(2)$         &     $36.7(2)$      &    $23.4(1)$  \\
   $128$   &     $42.3(2)$         &     $37.0(2)$      &    $23.8(1)$  \\
   $256$   &     $\mathbf{42.2(2)}$         &     $\mathbf{37.0(2)}$      &    $\mathbf{23.7(1)}$  \\
\hline
 \rowtitle{Variational\cite{PartridgePES}}        &     $42.37$    &       $37.14$        &   $23.79$       \\
\hline
 \rowtitle{Experiment\cite{Hall1967H20_rotspec}}    &   $42.36$    &     $37.14$    &   $23.79$   \\
\hline
 \rowtitle{Rigid Rotor}     &     $41.97$    &       $36.91$        &   $24.09$       \\
\end{tabular*}
\end{ruledtabular}
}
\end{table}

In summary, we have shown that the present approach can be used to obtain accurate energy differences both within a given $J$-manifold and between different manifolds.
Although the rotational energy levels of \ce{H2O} can be easily computed by other methods, the present results provide a clear proof of principle that the generalized PIMD formalism correctly projects onto specified rotational manifolds.
This establishes the basis for the calculation of rotationally resolved tunneling splittings. %

\subsection{Ammonia} \label{sec:NH3}

Ammonia, \ce{NH3}, is the prototypical system exhibiting a tunneling splitting associated with its umbrella inversion motion.
It is a symmetric-top molecule, and its rotation--inversion levels are labeled by $J_K^\pm$, where $J$ is the total angular-momentum quantum number, $K$ is its projection onto the molecular axis, and $\pm$ denotes the inversion parity.\cite{GordyCookBook}
Unlike $J$ and the parity, however, $K$ is not a good quantum number.
As such, it is not possible to use symmetry projection within the present PIMD framework to isolate states with a specified value of $K$.
Nonetheless, we can label the states by $K$, which, together with the parity, %
are incorporated within the index $n$
introduced in Sec.~\ref{sec:Theory}.

\begin{table}[htbp]
\caption{Character table for the $D_\mathrm{3h}$ molecular symmetry group of ammonia. Row headings denote conventional irrep labels and column headings correspond to conjugacy classes with the order of the class indicated by $g_\alpha$.}
\label{tab:char_tab_D3h}
{\renewcommand{\arraystretch}{1.5}
\begin{ruledtabular}
\begin{tabular*}{\columnwidth}{@{\extracolsep{\fill}}lcccccc}
$D_\mathrm{3h}$& $E$ & \phantom{+}$(123)$ & \phantom{+}$(12)$ & \phantom{+}$E^*$ & \phantom{+}$(123)^*$ & \phantom{+}$(12)^*$ \\
$g_\alpha$  & 1   & \phantom{+}2       & \phantom{+}3       & \phantom{+}1     & \phantom{+}2         & \phantom{+}3         \\
\hline
$A_1'$ & $1$ & $\phantom{-}1$ & $\phantom{-}1$ & $\phantom{-}1$ & $\phantom{-}1$ & $\phantom{-}1$\\
$A_1''$ & $1$ & $\phantom{-}1$ & $\phantom{-}1$ & $-1$ & $-1$ & $-1$\\
$A_2'$ & $1$ & $\phantom{-}1$ & $-1$ & $\phantom{-}1$ & $\phantom{-}1$ & $-1$ \\
$A_2''$ & $1$ & $\phantom{-}1$ & $-1$ & $-1$ & $-1$  & $\phantom{-}1$\\
$E'$ & $2$ & $-1$ & $\phantom{-}0$ & $\phantom{-}2$ & $-1$ & $\phantom{-}0$\\
$E''$ & $2$ & $-1$ & $\phantom{-}0$ & $-2$ & $\phantom{-}1$ & $\phantom{-}0$\\
\end{tabular*}
\end{ruledtabular}
}
\end{table}

\begin{figure}[htbp]
    \centering
    \includegraphics[width=0.95\linewidth]{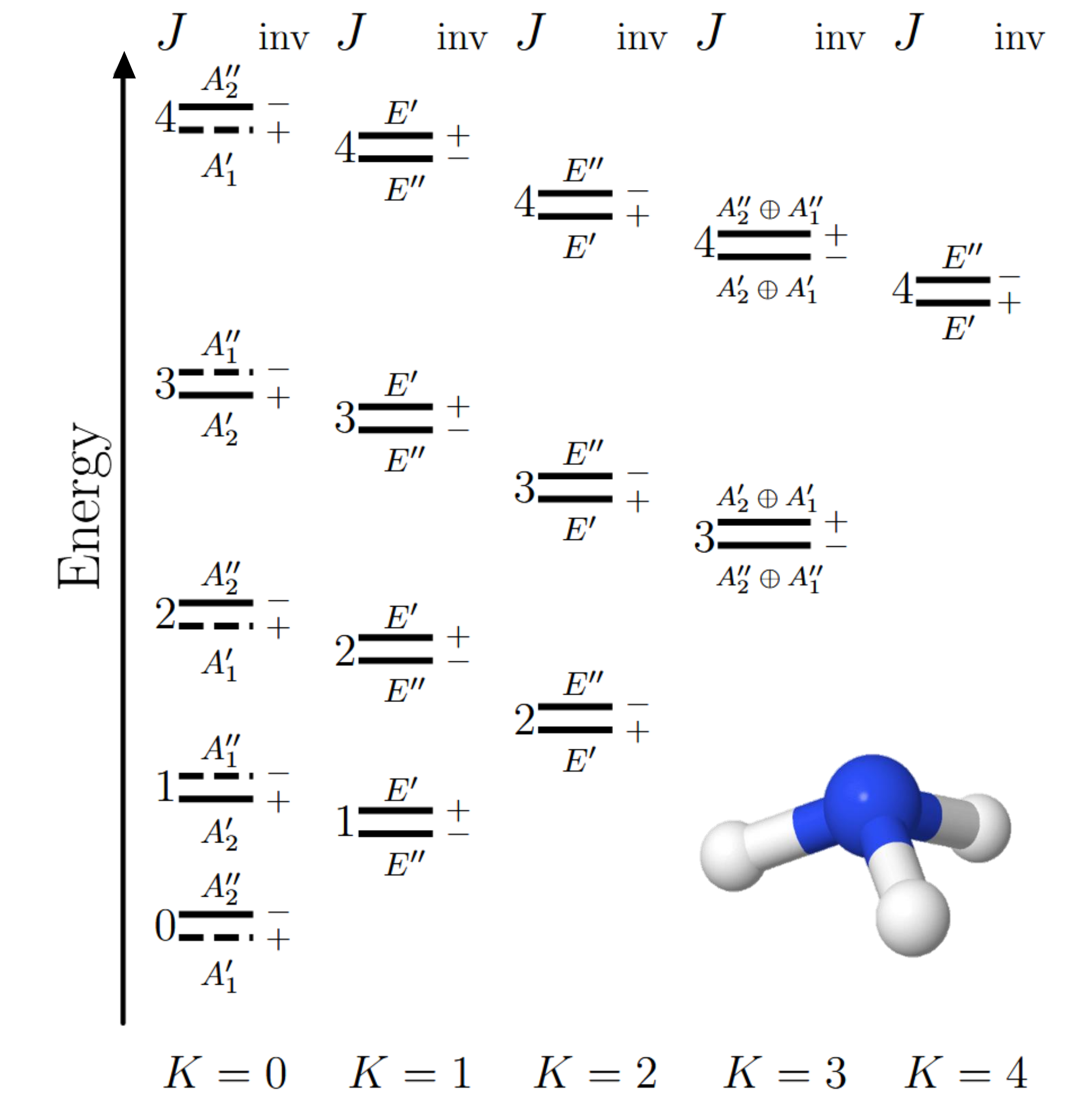}
    \caption{Rotation--inversion energy levels of ammonia, labeled by their irreps. Dashed lines denote states forbidden by nuclear spin statistics. For each tunneling doublet, the quantum number $J$ and the inversion parity $\pm$ are shown on the left and right, respectively. The $K$ value corresponding to each column is indicated at the bottom.
    The equilibrium structure of the ammonia molecule is shown in the lower right corner. %
    }
    \label{fig:ammonia}
\end{figure}

The molecular symmetry group of ammonia is isomorphic to $D_{3\mathrm{h}}$, whose character table is presented in \Tab{tab:char_tab_D3h}\@.
In addition to the pair permutation operators already encountered in the case of water, one must now also consider the cyclic hydrogen permutation operators, such as $(123)$. %
The relevant rotation--inversion levels, labeled by their irreps, are shown in Fig.~\ref{fig:ammonia}, where dashed lines indicate states forbidden by nuclear spin statistics.\cite{Bunker,BunkerBook}

In the present work, we use the PIMD method to evaluate the tunneling splittings for the $J_{K=0}$ levels with $J=0,\dots,4$, as well as for $J_K=1_1$.
These states are selected as they can be isolated on symmetry grounds.
By contrast, for example, the splittings between the $E'$ and $E''$ states for $J\ge2$ cannot be extracted within the present framework, as multiple states of $E'$ and $E''$ symmetry %
contribute simultaneously to the relevant symmetrized partition functions.

Although Eq.~(\ref{eq:genTunSplit}) can be applied directly to obtain the rotationally resolved tunneling splittings, such a treatment requires the evaluation of at least 5 distinct ratios among the symmetrized partition functions, because the $D_\mathrm{3h}$ group has 6 conjugacy classes.
However, for certain rotation--inversion manifolds, especially at low $J$, only a subset of the irreps of $D_\mathrm{3h}$ is present. As a result, the number of symmetrized partition functions required to isolate a given splitting is reduced.
It is therefore advantageous to restrict the analysis to the smaller symmetry subspace relevant to the manifold under consideration.
For example, for the rotational ground state of ammonia, only two levels with $A_1'$ and $A''_2$ symmetry are present, which is equivalent to reducing the symmetry from $D_\mathrm{3h}$ to $C_\mathrm{s}$.
In this case, the ground-state tunneling splitting can be written in the simplified form
\begin{equation} \label{Delta00}
    \Delta(0_0) = \frac{1}{\beta} \ln \left( \frac{Z_P^{(J=0)}+Z_{P'}^{(J=0)}}{Z_P^{(J=0)}-Z_{P'}^{(J=0)}}\right),
\end{equation}
where $\hat{P}$ can be chosen as one of $E$, $(123)$, and $(12)^*$, and $\hat{P}'$ can be chosen as one of $(12)$, $E^*$ or $(123)^*$.

More generally, for any $J_{K=0}$ level, the tunneling splittings can be expressed in a specific form of Eq.~\eqref{Delta00}:
\begin{equation} \label{DeltaJ0}
    \Delta(J_{K=0}) = \frac{1}{\beta} \ln \left( \frac{Z_{(12)^*}^{(J)}+Z_{(12)}^{(J)}}{Z_{(12)^*}^{(J)}-Z_{(12)}^{(J)}}\right).
\end{equation}
For $J=1$ and 2, this follows straightforwardly from the fact that the characters of $(12)$ and $(12)^*$ vanish in the $E'$ and $E''$ irreps, so that only the $K=0$ tunneling doublet contributes to  Eq.~\eqref{DeltaJ0}, which  directly yields the corresponding splitting.
For $J\ge3$, the situation is more subtle.
Since the $K=3$ states (and $K=6,9,\dots$) %
contain levels with $A_1'$, $A_2''$, $A_1''$, and $A_2'$ symmetry,\cite{Bunker,BunkerBook}
the $K=0$ doublet can no longer be isolated by symmetry projection alone.
Nevertheless, Eq.~\eqref{DeltaJ0} remains valid.
The reason is that the $K=3,6,9,\dots$ states form accidentally degenerate pairs whose characters have opposite signs under the operations $(12)$ and $(12)^*$.
As such, their contributions to $Z_{(12)}^{(J)}$ and $Z_{(12)^*}^{(J)}$ cancel with each other.
Strictly speaking, this degeneracy is not exact, since these accidental pairs are split by very weak higher-order rovibrational couplings.
However, such couplings are far smaller than the energy scales relevant to the present calculations and can therefore be safely neglected here.

A simplified expression for the splitting of $J_K=1_1$ levels can be obtained by noting that no states within the $J=1$ manifold belong to the $A_1'$ and $A_2''$ irreps.
We can therefore write
\begin{equation} \label{Delta11}
    \Delta(1_1) = \frac{1}{\beta} \mathrm{ln} \left( \frac{Z_E^{(J=1)}+2Z_{(12)}^{(J=1)}+2Z_{(123)^*}^{(J=1)}+Z_{(12)^*}^{(J=1)}}{Z_E^{(J=1)}-2Z_{(12)}^{(J=1)}-2Z_{(123)^*}^{(J=1)}+Z_{(12)^*}^{(J=1)}}\right) .
\end{equation}

Equations~\eqref{Delta00}--\eqref{Delta11} show that the evaluation of the rotationally resolved tunneling splittings reduces to only a small number of partition-function ratios. %
For the $J_{K=0}$ levels with $J=0,\ldots,4$, Eqs.~\eqref{Delta00}--\eqref{DeltaJ0} can be obtained from only a single partition-function ratio, $Z_{(12)}^{(J)}/Z_{(12)^*}^{(J)}$.
However, the $J_K=1_1$ splitting [Eq.~\eqref{Delta11}] involves four symmetrized partition functions.
Using $Z_E^{(J=1)}$ as the reference, this splitting can be evaluated from the three ratios relative to $Z_{(12)}^{(J=1)}$, $Z_{(123)^*}^{(J=1)}$, and $Z_{(12)^*}^{(J=1)}$. %

The PIMD simulations were performed on the \mbox{AMMPOT4} global analytical PES developed by Marquardt et al.\cite{MarquardtNH3}
The required free-energy differences were evaluated by TI\@.
Each TI calculation employed 96 independent parallel trajectories and was performed sequentially over the quadrature points.
At the initial quadrature point, the trajectories were thermalized for $5\,\mathrm{ps}$, whereas at each subsequent quadrature point, the simulation was initialized from the final configuration of the preceding point and propagated for $50\,\mathrm{fs}$ to relax to the new equilibrium distribution.\footnote{We found that a relaxation time of $50\,\mathrm{fs}$  is sufficient for ammonia in the present work, as evidenced by the convergence with respect to $n_\xi$ shown in \Tab{tab:Ammonia_res1} and \Tab{tab:Ammonia_res2}. A longer relaxation time may be required for other molecular systems, as suggested by previous studies on hydronium, methanol, and malonaldehyde.\cite{PIMDtunnel,malonaldehydePIMD}}
Data were collected over $25\,\mathrm{ps}$ at each quadrature point.

All simulations were carried out at $T=100$\,K\@.
At this temperature, contamination from vibrational excitation is negligible.
The lowest vibrationally excited tunneling doublet has its lower component at $932\,\mathrm{cm}^{-1}$, and its upper component lies about $36\,\mathrm{cm}^{-1}$ higher.\cite{URBAN1999highResNH3}
A simple four-state model similar to that used in \Refx{malonaldehydePIMD} suggests that the contribution to the predicted ground-state tunneling splitting will only be about $0.0001\,\mathrm{cm}^{-1}$, which is much smaller than the statistical uncertainties of the current calculations.

\begin{table*}[htbp]
\caption{Tunneling splittings [in $\mathrm{cm}^{-1}$] for different rotational levels $J_{K=0}$ of \ce{NH3} calculated by PIMD using Eq.~\eqref{DeltaJ0} in terms of $Z_{(12)}^{(J)}/Z_{(12)^*}^{(J)}$ ratios from simulations at 100\,K with various parameters. %
The boldfaced values obtained with $N=256$ and $n_\xi=30$ are taken as the final PIMD results.}%
\label{tab:Ammonia_res1}
{\renewcommand{\arraystretch}{1.5}
\begin{ruledtabular}
\begin{tabular*}{\textwidth}{@{\hspace{4pt}}@{\extracolsep{\fill}}ccccccc}
 $N$  & $n_\xi$ & \head{$\Delta(0_0)$} & \head{$\Delta(1_0)$} & \head{$\Delta(2_0)$} & \head{$\Delta(3_0)$} & \head{$\Delta(4_0)$} \\
\hline
 64  & 10 &    $0.809(9)$   &    $0.797(9)$   &    $0.77(1)$   &    $0.73(2)$   &    $0.69(4)$ \\
 64  & 15 &    $0.807(8)$   &    $0.795(8)$   &    $0.78(1)$   &    $0.76(2)$   &    $0.76(4)$ \\
 128 & 10 &    $0.76(1)$    &    $0.75(1)$    &    $0.74(1)$   &    $0.72(2)$   &    $0.72(4)$ \\
 128 & 15 &    $0.758(7)$   &    $0.751(7)$   &    $0.738(9)$  &    $0.73(2)$   &    $0.74(4)$ \\
 128 & 20 &    $0.761(7)$   &    $0.752(7)$   &    $0.74(1)$   &    $0.71(2)$   &    $0.69(4)$ \\
 128 & 25 &    $0.759(6)$   &    $0.752(7)$   &    $0.738(9)$  &    $0.72(2)$   &    $0.70(4)$ \\
 256 & 15 &    $0.77(1)$    &    $0.76(1)$    &    $0.74(1)$   &    $0.69(2)$   &    $0.62(4)$ \\
 256 & 20 &    $0.752(8)$   &    $0.741(8)$   &    $0.72(1)$   &    $0.70(2)$   &    $0.69(4)$ \\
 256 & 25 &    $0.758(8)$   &    $0.750(8)$   &    $0.74(1)$   &    $0.71(2)$   &    $0.68(4)$\\
 256 & 30 &    $\mathbf{0.748(6)}$   &    $\mathbf{0.741(6)}$   &    $\mathbf{0.726(9)}$   &    $\mathbf{0.70(2)}$   &    $\mathbf{0.66(4)}$\\
\hline
\multicolumn{2}{l}{Variational\cite{Wichmann-MP-2020-1752946}} & $0.751$ & $0.742$ & $0.724$ & $0.698$ & $0.665$\\
\hline
\multicolumn{2}{l}{Experiment\cite{Urban1984_NH3}} & $0.793$ & $0.783$ & $0.764$ & $0.735$ & $0.699$\\
\end{tabular*}
\end{ruledtabular}
}
\end{table*}

\Tab{tab:Ammonia_res1} presents the tunneling splittings $\Delta(J_{K=0})$ for $J=0,\ldots,4$ obtained from Eq.~\eqref{DeltaJ0} in terms of $Z_{(12)}^{(J)}/Z_{(12)^*}^{(J)}$ ratios with different bead numbers, $N$, and numbers of TI quadrature points, $n_\xi$.
For each value of $N$, the results obtained with the two largest values of $n_\xi$ are consistent with each other within statistical uncertainty, %
indicating convergence with respect to the TI quadrature.
Convergence with respect to $N$ can be assessed by comparing the results obtained with the largest $n_\xi$ at each bead number.
The results change significantly as $N$ increases from 64 to 128, whereas those obtained with $N=128$ and 256 are statistically consistent, indicating that convergence with respect to $N$ has been reached. %
We therefore take the results obtained with the largest parameter set ($N=256$ and $n_\xi=30)$ as the final PIMD values.

At the bottom of \Tab{tab:Ammonia_res1}, the final PIMD results are compared with the numerically exact values from variational wavefunction calculations\cite{Wichmann-MP-2020-1752946} (on the same PES) and with experiment.\cite{Urban1984_NH3}
As can be seen, the PIMD results are consistent with the variational values within statistical uncertainty. 
This level of agreement is expected, since once all numerical parameters are converged, our approach yields the numerically exact tunneling splittings of the underlying PES up to statistical error. 

In reality, one component of each $J_{K=0}^{\pm}$ tunneling doublet is absent from the spectrum due to nuclear-spin statistical selection rules.
Accordingly, these splittings are not directly accessible by experiments.
They can nevertheless be inferred by fitting the allowed transitions to an effective spectroscopic Hamiltonian and extracting the splitting that would be obtained in the absence of the nuclear-spin restriction.
As seen from \Tab{tab:Ammonia_res1}, both the PIMD and variational results based on the present PES underestimate all five experimental splittings.
Since both theoretical treatments are numerically exact, this discrepancy reflects imperfections in the underlying PES\@.\footnote{Note that other PESs have also been constructed and successfully used to obtain energy levels in good agreement with experiment [Yurchenko et al.\ J. Phys. Chem. A 113, 11845–11855 (2009)]. %
However, the energies of the spin-forbidden levels required to obtain the $J_{K=0}$ splittings were not published.}
Nevertheless, the theoretical values reproduce the observed decrease of the tunneling splitting with increasing $J$.

\begin{figure}[htbp]
    \centering
    \includegraphics[width=0.95\linewidth]{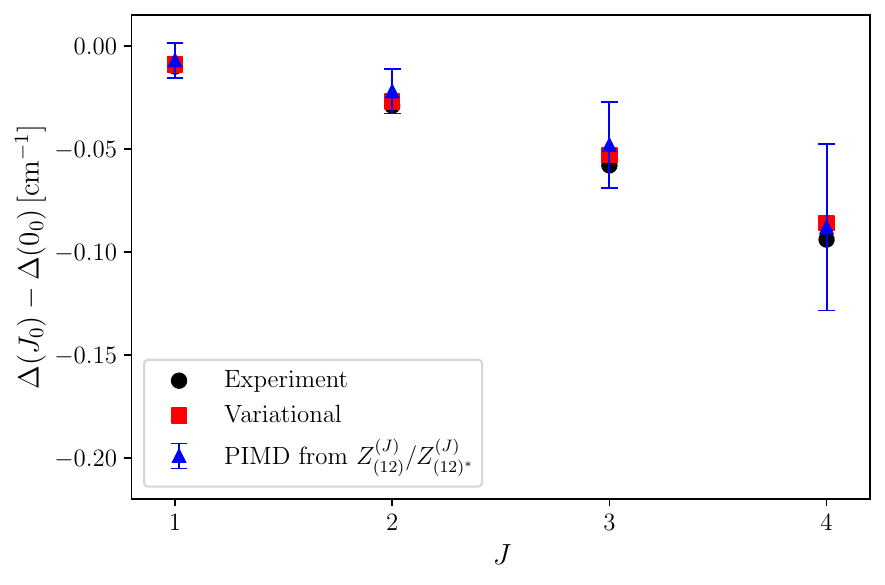}
    \caption{Difference between the %
    rotationally excited and ground-state tunneling splittings
    as a function of the total angular-momentum quantum number $J$.
    Data are presented for the final converged PIMD results, the variational wavefunction results, and the experimental values.}
    \label{fig:Splitting_diff_vs_J}
\end{figure}

To show this trend more clearly, Fig.~\ref{fig:Splitting_diff_vs_J} plots, for each $J$, the difference between $\Delta(J_{K=0})$ and the ground-state splitting $\Delta(0_0)$.
The theoretical and experimental curves are seen to be in close agreement.
This indicates that the present PES underestimates the tunneling splittings primarily in an overall sense, while describing the relative variation between different rotational levels rather accurately.

We observe that the statistical uncertainties increase with the total angular-momentum quantum number $J$.
One reason is that, according to Eq.~\eqref{eq:TrD}, the trace of the Wigner $D$-matrix becomes increasingly oscillatory as $J$ increases, which reduces the sampling efficiency.
A second reason is that, for $J>0$, this trace  depends on the relative orientation of the first and last beads of a given ring-polymer configuration.
In the present PIMD simulations, changes in this relative orientation arise from slow collective motions of the entire polymer. %
This further reduces the efficiency of the sampling.
We also find that this difficulty becomes more pronounced at lower temperatures.
In principle, the sampling of the relative end-to-end orientation could be improved by introducing suitable global Monte Carlo updates.
One possible example would be an inter-bead rotational update accepted according to a Metropolis criterion.
Such developments are currently in progress.\cite{Yuchen-PIHMC-2026}

\begin{table}[htbp]
\caption{Tunneling splittings [in $\mathrm{cm}^{-1}$] for $J_K=0_0$ and $1_1$ of \ce{NH3} calculated by PIMD using Eqs.~\eqref{Delta00} and \eqref{Delta11} from simulations at 100\,K with various parameters. %
The boldfaced values obtained with $N=256$ and $n_\xi=30$ are taken as the final PIMD results. %
}
\label{tab:Ammonia_res2}
{\renewcommand{\arraystretch}{1.5}
\begin{ruledtabular}
\begin{tabular*}{\columnwidth}{@{\hspace{3pt}}@{\extracolsep{\fill}}ccccc}
 $N$ & $n_\xi$ & \head{$\Delta(0_0)$\footnotemark[1]} & \head{$\Delta(0_0)$\footnotemark[2]}& \head{$\Delta(1_1)$} \\
\hline
 64   & 10 &   0.819(8)    &    0.813(9)    &   $0.80(2)$ \\
 64   & 15 &   0.805(7)    &    0.808(8)    &   $0.81(2)$ \\
 128  & 10 &   0.76(1)     &    0.75(1)     &   $0.75(3)$ \\
 128  & 15 &   0.776(9)    &    0.779(8)    &   $0.78(2)$ \\
 128  & 20 &   0.768(7)    &    0.776(7)    &   $0.78(2)$ \\
 128  & 25 &   0.768(6)    &    0.756(7)    &   $0.74(1)$ \\
 256  & 15 &   0.74(1)     &    0.77(1)     &   $0.78(2)$ \\
 256  & 20 &   0.767(8)    &    0.776(8)    &   $0.78(2)$ \\
 256  & 25 &   0.747(7)    &    0.754(6)    &   $0.77(1)$ \\
 256  & 30 &   $\mathbf{0.758(7)}$    &    $\mathbf{0.760(8)}$    &   $\mathbf{0.75(2)}$ \\ 
\hline
\multicolumn{2}{l}{Variational\cite{Wichmann-MP-2020-1752946}} & $0.751$ & $0.751$ &  $0.748$ \\
\hline
\multicolumn{2}{l}{Experiment\cite{Urban1984_NH3}}       & $0.793$ & $0.793$ &  $0.790$ \\
\end{tabular*}
\end{ruledtabular}
}
\footnotetext[1]{Splittings obtained from Eq.~\eqref{Delta00} with $Z_{(12)}^{(J=0)}/Z_E^{(J=0)}$}
\footnotetext[2]{Splittings obtained from Eq.~\eqref{Delta00} with $Z_{(123)^*}^{(J=0)}/Z_E^{(J=0)}$}
\end{table}
\Tab{tab:Ammonia_res2} presents the $J_K=1_1$ tunneling splitting,
which is less strongly perturbed by the rotation than the $J_K=1_0$ splitting.
The PIMD result is obtained from Eq.~\eqref{Delta11}.
As in the previous cases, convergence is reached at $N=256$ and $n_\xi=30$, which is therefore taken as the final PIMD value.
The converged result %
is in good agreement with the variational benchmark within statistical uncertainty.

In addition, \Tab{tab:Ammonia_res2} also includes the ground-state tunneling splitting $\Delta(0_0)$ evaluated from Eq.~\eqref{Delta00}, with $\hat{P}=E$ and $\hat{P}'=(12)$ or $(123)^*$, instead of Eq.~\eqref{DeltaJ0}.
These alternative evaluations of $\Delta(0_0)$ are likewise consistent with the exact value %
and exhibit a statistical uncertainty comparable to that obtained previously.
This demonstrates that (once converged) the various alternative formulations of the splitting are equivalent, and suggests that there are no particular advantages or disadvantages to the various formulations beyond reducing the number of ratios required.

\section{Conclusion}\label{sec:conclusion}
In this paper, we have extended the PIMD method developed in \Refx{PIMDtunnel} to include a projection onto the manifold of states with a selected angular-momentum quantum number $J$.  Together with a projection onto selected discrete-symmetry subspaces of the molecular system, this enables the exact calculation of tunneling splittings in rotationally excited states.
In practice, the tunneling splittings are obtained from ratios of symmetrized partition functions corresponding to a set of permutation–inversion operations, and these ratios are factorized into a free-energy contribution and a $J$-dependent prefactor.
In this way, tunneling splittings for a range of values of $J$ can be obtained simultaneously by post-processing data gathered from a set of PIMD simulations. %

The method was first applied to \ce{H2O} to compute the energy differences between states in the first two rotational manifolds.
This validated our formalism and also demonstrated that the method is not limited to tunneling splittings but can also be used to calculate rotational levels beyond the rigid-rotor approximation. %
It was then applied to \ce{NH3} to evaluate rotationally resolved tunneling splittings for several low-lying rotational states.
For this system, the PIMD results were found to be in excellent agreement with the numerically exact variational wavefunction results on the same PES\@.
Comparison with experiment further showed that, although the present PES slightly underestimates the absolute splittings, the dependence of the splittings on $J$ is well described.

Although the present applications focused on \ce{H2O} and \ce{NH3}, for which exact variational calculations remain straightforward and inexpensive, the main advantage of the PIMD approach lies in its more favorable scaling with system size.
In addition, the method is formulated directly in Cartesian coordinates and is readily applicable to other molecular systems, including extremely floppy molecules exhibiting strong correlation, for which traditional approaches face great challenges.\cite{hinkle_diffusion_2011,schmiedt_symmetry_2015,wodraszka_ch5_2015}
Therefore, the present work provides a practical alternative route to the rigorous calculation of rotationally resolved tunneling splittings, including larger or highly floppy molecular systems.

In principle, the PIMD method can also be extended to excited vibrational states, provided that the target state is the lowest-lying state within the chosen symmetry specified by both the irrep and the value of $J$.
For example, the fundamental excitation of the $B_2$ asymmetric stretch of \ce{H2O} within the $J=0$ manifold could in principle be obtained from the present formalism.
In practice, however, the relevant expressions involve sums and differences of $Z_P^{(J=0)}$ that are nearly equal to each other [Eq.~\eqref{ZsJ0}], which would lead to large statistical uncertainties.
It may nevertheless be possible to obtain a reliable result at a higher temperature.
Another potential candidate is the formic acid dimer, for which vibrationally excited tunneling splittings have been probed with infrared spectroscopy.\cite{Birer2009review}

While the present work employed TI for the free-energy calculations, the formalism itself is not restricted to this approach, and one direction of our ongoing work is the development of more efficient free-energy sampling methods for tunneling splitting calculations.\cite{Yuchen-PIHMC-2026}
This would further enhance the capability of the PIMD method to address challenging rovibrational problems beyond the reach of conventional approaches.

Finally, the rotational-projection formalism developed here may also provide inspiration for other approaches,
such as an improved semiclassical instanton theory for rotationally excited states.
This approach would be more efficient than the PIMD method, as it does not require sampling, although it does introduce an approximation.
Nonetheless, the instanton approach can provide accurate results, especially when combined with perturbative corrections,\cite{AnharmInst} as long as the anharmonicity is not too strong.
However, for very floppy molecules, the PIMD method will remain the method of choice.

\section*{Supplementary Material}
See the supplementary material for raw data %
used for calculating the tunneling splittings, %
including TI plots for ammonia, and further details on how error propagation was used. %

\section*{Acknowledgements}
The authors thank George Trenins and Hannes Hoppe for helpful discussions.
The authors acknowledge financial support from the Swiss National Science Foundation through the project
titled ‘Nonadiabatic effects in chemical reactions’ (Grant Number – 207772).

\section*{Author contributions}
\noindent \textbf{L\'{e}a Zupan}: Methodology (lead); Investigation (lead); Software (lead); Formal analysis (equal); Visualization (lead); Writing – original draft (lead).
\\
\textbf{Yu-Chen Wang}: Methodology (supporting); Investigation (supporting); Software (supporting); Formal analysis (equal); Supervision (equal); Writing – original draft (supporting); Writing – review \& editing (equal).
\\
\textbf{Jeremy O. Richardson}: Conceptualization (lead); Methodology (supporting); Formal analysis (supporting); Supervision (equal); Writing – original draft (supporting); Writing – review \& editing (equal).

\section*{Data Availability}
The data that support the findings of this study are available within the article and its supplementary material.

\appendix

\section{Forces associated with the Eckart spring} \label{app:EckartForce}
\subsection{Eckart spring with the identity}
The Eckart spring without permutation--inversion operations (or equivalently, with the identity $\hat{P}=E$) has the following form:
\begin{equation}
      U_{E}(\boldsymbol{r};\widetilde{\boldsymbol{\Omega}}) 
      =\sum^{n_\mathrm{at}}_{a=1}\frac{m_a \omega_N^2}{2}
      \left\lVert \vec{r}_a^{(N)} - \widetilde{\vec{r}}_a^{(1)} \right\rVert^2,
\end{equation}
where we recall that  $\widetilde{\vec{r}}^{(1)} = \hat{R}(\widetilde{\boldsymbol{\Omega}}) \vec{r}^{(1)}$ is the rotated configuration that minimizes the Eckart spring. The Eckart force acting on a single atom $b$ in the $i$-th bead is calculated through:\cite{PIMDtunnel}
\begin{equation}
\begin{split}
    \mathbf{F}^{(i)}_{E,b}
    &=-\frac{\partial U_E(\boldsymbol{r};\widetilde{\boldsymbol{\Omega}})}{\partial\vec{r}_b^{(i)}} %
    = -\frac{\omega_N^2}{2}\sum_{\mu,\nu=0}^{3}q_{\mu}^{(0)}\frac{\partial C_{\mu\nu}}{\partial \vec{r}_b^{(i)}}q_{\nu}^{(0)} ,
\end{split}
\end{equation}
where the $4 \times 4$ real symmetric matrix, $C_{\mu\nu}$, and its normalized eigenvector $q_\mu^{(0)}$ (corresponding to the smallest eigenvalue) are obtained from the quaternion algorithm used to find the optimal rotation matrix $\hat{R}(\widetilde{\boldsymbol{\Omega}})$ as detailed in \Refx{Krasnoshchekov2014Eckart}.
Note that as $C_{\mu\nu}$ only depends on the first and last beads, this force is zero unless $i\in\{1,N\}$. 

\subsection{Eckart spring with permutation}
If $\hat{P}$ is a symmetry operation other than identity, the end-bead configuration entering the Eckart spring is not $\vec{r}^{(N)}$ itself, but $\hat{P}^{-1}\vec{r}^{(N)}$. Since a permutation of two or more identical atoms in bead $N$ does not change the physical configuration of that bead but only relabels the atomic coordinates, the corresponding Eckart-spring forces are obtained by a simple relabeling of the forces in the non-permuted case. Thus, if $\hat{P}^{-1}$ exchanges identical atoms $a$ and $b$ [e.g., $\hat{P}^{-1}=(ab)$], the force acting on atom $a$ of bead $N$ is
\begin{equation}
    \mathbf{F}^{(N)}_{P,a} = \mathbf{F}^{(N)}_{E,b}.
\end{equation}
That is, the force acting on atom $a$ is equal to the non-permuted Eckart-spring force acting on the atom to which $a$ is mapped by $\hat{P}^{-1}$.

\subsection{Eckart spring with inversion}
Because of the nature of the molecules studied, Refs.\ \onlinecite{PIMDtunnel} and \onlinecite{malonaldehydePIMD} only needed to consider pure permutation operations. In the present work, however, pure inversion and more general permutation--inversion operations are also required to treat \ce{H2O} and \ce{NH3}.

For a pure inversion operation, $\hat{P}=E^*$, the inverse operation, $\hat{P}^{-1}=E^*$, is identical to the operation itself.
The coordinates of atom $a$ in bead $N$ are therefore inverted through the center of mass according to
\begin{equation}
    \hat{P}^{-1} \vec{r}_a^{(N)} = -\big(\vec{r}_a^{(N)} - \vec{r}_{\mathrm{com}}^{(N)}\big) + \vec{r}_{\mathrm{com}}^{(N)},
\label{eq:Inversion}
\end{equation}
where
\begin{equation}
    \vec{r}_{\mathrm{com}}^{(N)} = \sum_{b=1}^{n_\mathrm{at}}\frac{m_b}{M}\vec{r}_b^{(N)}
\end{equation}
and $M=\sum_{b=1}^{n_\mathrm{at}} m_b$ is the total mass of the molecular system.
Unlike a pure permutation, this operation changes the configuration of bead $N$.
The corresponding force acting on atom $a$ is therefore obtained by applying the chain rule to Eq.~\eqref{eq:Inversion}, which yields
\begin{equation}\label{eq:force_inversion}
    \mathbf{F}^{(N)}_{E^*,a} = \frac{2m_a}{M}
    \sum_{b=1}^{n_\mathrm{at}}\mathbf{F}^{(N)}_{E,b} - \mathbf{F}^{(N)}_{E,a},
\end{equation}
where the forces on the right-hand side are the non-permuted Eckart-spring forces evaluated for the inverted geometry $\hat{P}^{-1} \vec{r}^{(N)}$.

For a general permutation--inversion operator, the required transformation is obtained by replacing the forces on the right-hand side of Eq.~\eqref{eq:force_inversion} by those associated with the corresponding permutation operator, again evaluated for the inverted geometry. %

\bibliography{new_refs,references}
\end{document}


\begin{CJK*}{UTF8}{gbsn}

\title{Supplementary material: Exact tunneling splittings of rotationally excited states from symmetrized path-integral molecular dynamics}

\author{L\'{e}a Zupan}
\altaffiliation{On exchange from \'Ecole Polytechnique F\'ed\'erale de Lausanne (EPFL)} %
\author{Yu-Chen Wang (汪宇晨)}
\email{wangyuc@phys.chem.ethz.ch}
\author{Jeremy O. Richardson}
\email{jeremy.richardson@phys.chem.ethz.ch}
\affiliation{\mbox{Institute of Molecular Physical Science, ETH Z\"{u}rich, 8093 Z\"{u}rich, Switzerland}}

\date{\today}

\maketitle
\end{CJK*}

\noindent This supplementary material provides the raw data calculated for water and ammonia, which are combined to provide the predicted tunneling splittings, as explained in the main text.
In each case, statistical errors obtained from bootstrapping are reported in parentheses as $1\sigma$.

\section{Water}

\begin{table}[H]
\centering
\caption{ Prefactors $\langle u_P^{(J)}\rangle \times 10^{4}$ for $J=0$ and 1 obtained from different simulation parameters sets.}
\begin{tabular*}{\textwidth}{@{\extracolsep{\fill}}cccd{1.7}d{2.7}d{2.7}d{2.7}d{2.7}d{2.7}d{2.7}d{2.7}}
\toprule
$T\,[\mathrm{K}]$ & $N$ & $\Delta t$\,[fs] & \head{$\langle u_E^{(J=0)}\rangle$} & \head{$\langle u_{(12)}^{(J=0)}\rangle$} &  \head{$\langle u_{E^*}^{(J=0)}\rangle $} & \head{$\langle u_{(12)^*}^{(J=0)}\rangle$} 
 & \head{$\langle u_E^{(J=1)}\rangle$} & \head{$\langle u_{(12)}^{(J=1)}\rangle$} &  \head{$\langle u_{E^*}^{(J=1)}\rangle $} & \head{$\langle u_{(12)^*}^{(J=1)}\rangle$} \\
\midrule

\multirow{5.5}{*}{100} 
  &         32            & 0.2   & 3.2159(2) & 3.2160(2) & 3.2160(2) & 3.2159(2) & 5.973(14) & -2.166(6) & -2.441(4) & -1.3716(12) \\
\cmidrule{2-11}
  &         64            & 0.2   & 1.12517(6) & 1.12527(6) & 1.12521(8) & 1.12519(8) & 2.081(4) & -0.756(2) & -0.850(2) &  -0.478(4) \\
 
\cmidrule{2-11}
  &     128  & 0.2 & 0.39577(2) & 0.39578(2)  & 0.39578(2)  & 0.39577(2)  & 0.7318(16) & -0.2644(8) & -0.2981(6) & -0.1665(14) \\
\cmidrule{2-11}
  &    256   & 0.2 & 0.13960(2) & 0.13960(2) & 0.13961(2) & 0.13960(2) & 0.2570(6) & -0.0934(2) & -0.1051(2) & -0.0588(4) \\
\bottomrule
\end{tabular*}\label{tab:SI_Water_prefactors}
\end{table}

\begin{table}[H]
\centering
\setlength{\tabcolsep}{8pt}
\caption{ Partition-function ratios $Z_P^{(J)}/Z_E^{(J=0)}$ for $J=1$ and different $\hat{P}$ obtained from different simulation parameters sets. The ratios for $J=0$ are all strictly 1 in the zero-temperature limit.}
\begin{tabular}{ccccccc}
\toprule
$T\,[\mathrm{K}]$ & $N$ & $\Delta t$\,[fs] & $E$ & $(12)$ &  $(E^*)$ & $(12)^*$  \\
\midrule
\multirow{5.5}{*}{100} 
  &         32            & 0.2   & $1.858(4)$ & $-0.673(2)$ & $-0.759(1)$ & $-0.426(4)$  \\
\cmidrule{2-7}
  &         64            & 0.2   & $1.849(4)$ & $-0.672(2)$ & $-0.756(1)$ &  $-0.425(4)$ \\
 
\cmidrule{2-7}
  &    128    & 0.2 & $1.849(4)$ & $-0.668(2)$  & $-0.753(1)$ & $-0.421(3)$  \\
\cmidrule{2-7}
  &    256    & 0.2 & $1.841(4)$ & $-0.669(2)$ & $-0.753(1)$ & $-0.421(3)$  \\
\bottomrule
\end{tabular}\label{tab:SI_Water_Z}
\end{table}

\section{Ammonia}

\begin{table}[H]
\centering
\caption{ Prefactors $\langle u_P^{(J)} \rangle \times 10^{5}$ and $\Delta F\,[\mathrm{Hartree}]$ obtained from the simulation of $Z_{(12)}^{(J')}/Z_{E}^{(J)}$ for various simulation parameters.}
\begin{tabular}{ccccd{3.7}d{2.8}d{2.8}d{2.6}d{2.6}d{2.6}d{2.6}d{1.7}}
\toprule
$T\,[\mathrm{K}]$ & $N$ & $\Delta t$\,[fs] & $n_\xi$ & \head{$\Delta F$} & \head{$\langle u_E^{(J=0)} \rangle$} & \head{$\langle u_{(12)}^{(J=0)}\rangle$}& \head{$\langle u_{E}^{(J=1)}\rangle$}& \head{$\langle u_{(12)}^{(J=1)}\rangle$} & \head{$\langle u_{(12)}^{(J=2)}\rangle$} & \head{$\langle u_{(12)}^{(J=3)}\rangle$} & \head{$\langle u_{(12)}^{(J=4)}\rangle$}\\
\midrule

\multirow{11}{*}{100} 
  & \multirow{2}{*}{64}   & 0.2   & 10 & 0.1043(2) & 4.5736(4) & 4.6382(7) & 10.71(1) & -3.472(8) & 1.95(1) & -0.82(1) & 0.27(1)\\
  & & 0.2 & 15 & 0.1047(2) & 4.5736(4) & 4.6398(7) & 10.73(1) & -3.456(7) & 1.92(1) & -0.79(1) & 0.240(9)\\
\cmidrule{2-12}
  & \multirow{4}{*}{128}  & 0.2 & 10 & 0.2117(5) & 1.6077(2) & 1.6305(2) & 3.764(4) & -1.218(3) & 0.679(4) & -0.280(4) & 0.084(4)\\
  &  & 0.2 & 15 & 0.2109(5) & 1.6076(2) & 1.6308(2) & 3.766(4) & -1.215(3) & 0.673(5) & -0.277(5) & 0.085(4)\\
  &  & 0.2 & 20 & 0.2113(3) & 1.6074(2) & 1.6304(3) & 3.757(4) & -1.210(3) & 0.666(4) & -0.271(4) & 0.082(3) \\
  &  & 0.2 & 25 & 0.2113(3) & 1.6074(2) & 1.6309(3) & 3.761(4) & -1.211(2) & 0.668(4) & -0.273(4) & 0.082(3)\\
 
\cmidrule{2-12}
  & \multirow{4}{*}{256} & 0.2 & 15 & 0.425(1) & 0.56695(6) & 0.57492(9) & 1.326(1) & -0.429(1) & 0.239(2) & -0.100(2) & 0.031(1) \\
  & & 0.2& 20 & 0.4226(8) & 0.56691(7) & 0.57481(8) & 1.323(2) & -0.4267(9) & 0.235(2) & -0.095(2) & 0.028(1) \\
  & & 0.2& 25 & 0.4249(8) & 0.56687(7) & 0.57483(8) & 1.327(1) & -0.426(1) & 0.233(2) & -0.094(2) & 0.028(1) \\
  & & 0.2& 30 & 0.4236(7) & 0.56692(6) & 0.57480(9) & 1.327(2) & -0.4274(9) & 0.236(2) & -0.097(2) & 0.030(1) \\

\bottomrule
\end{tabular}\label{tab:Ammonia_12_ratios_2}
\end{table}

\begin{table}[H]
\centering
\caption{Partition-function ratios $Z_{(12)}^{(J')}/Z_{E}^{(J)} \times 10^3$ with  various values of $J$ and $J'$ obtained from different simulation parameters.}
\begin{tabular}{ccccd{3.5}d{2.5}d{2.5}d{2.5}d{2.7}d{2.6}}
\toprule
$T\,[\mathrm{K}]$ & $N$ & $\Delta t$\,[fs] & $n_\xi$ & \head{$Z_{(12)}^{(J=0)}/Z_{E}^{(J=0)}$} & \head{$Z_{(12)}^{(J=1)}/Z_{E}^{(J=0)}$} & \head{$Z_{(12)}^{(J=2)}/Z_{E}^{(J=0)}$}& \head{$Z_{(12)}^{(J=3)}/Z_{E}^{(J=0)}$}& \head{$Z_{(12)}^{(J=4)}/Z_{E}^{(J=0)}$} & \head{$Z_{(12)}^{(J=1)}/Z_{E}^{(J=1)}$}\\
\midrule

\multirow{11}{*}{100} 
  & \multirow{2}{*}{64}   & 0.2   & 10 & 5.89(6) & -4.41(4) & 2.47(3) & -1.05(2) & 0.34(1) & -1.88(2) \\
  & & 0.2 & 15 & 5.79(5) & -4.31(4)& 2.40(3) & -0.99(2) & 0.30(1) & -1.83(2)\\
\cmidrule{2-10}
  & \multirow{4}{*}{128}  & 0.2 & 10 & 5.47(7) & -4.09(6) & 2.28(4) & -0.94(2) & 0.28(1) & -1.75(3) \\
  &  & 0.2 & 15 & 5.58(6)& -4.16(5) & 2.30(3) & -0.95(2) & 0.29(1) & -1.77(2)\\
  &  & 0.2 & 20 & 5.53(5) & -4.10(4) & 2.26(2) & -0.92(2) & 0.28(1) & -1.75(2) \\
  &  & 0.2 & 25 & 5.53(5) & -4.11(4) & 2.26(2) & -0.92(2) & 0.28(1) & -1.76(2)\\
 
\cmidrule{2-10}
  & \multirow{4}{*}{256} & 0.2 & 15 & 5.34(7) & -3.99(5)& 2.22(3) & -0.93(2) &  0.29(1) & -1.70(2)\\
  & & 0.2& 20 & 5.52(6)& -4.10(4)& 2.26(3)& -0.91(2) & 0.27(1) & -1.76(2)\\
  & & 0.2& 25 & 5.37(5) & -3.97(4) & 2.18(3)& -0.88(2) & 0.26(1) & -1.70(2)\\
  & & 0.2& 30 & 5.45(5)& -4.06(4) & 2.24(3)& -0.92(2) & 0.29(1) & -1.73(2)\\

\bottomrule
\end{tabular}\label{tab:Ammonia_12_ratios}
\end{table}

\begin{figure}[H]
    \centering

    \begin{subfigure}{\textwidth}
        \centering
        \includegraphics[width=0.5\linewidth]{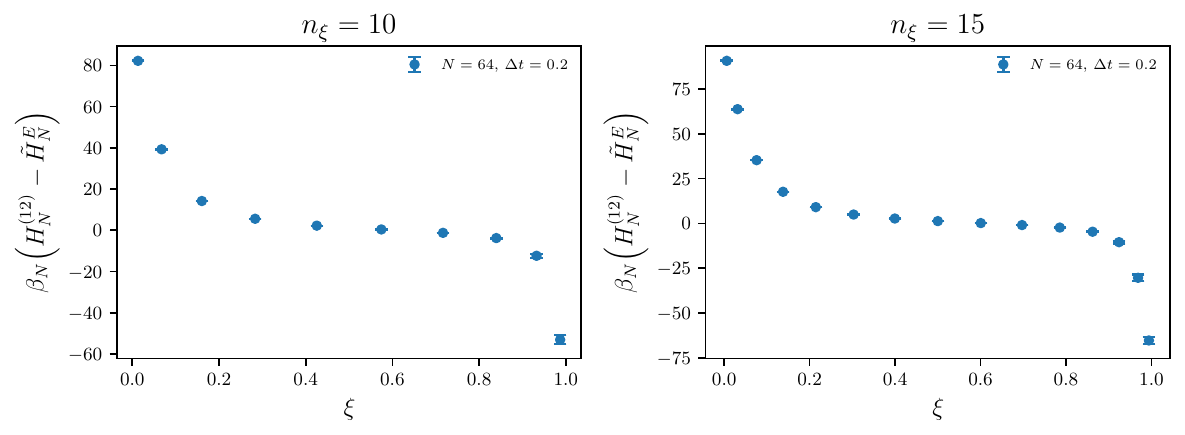}
        \caption{$\beta_N = 49$}
    \end{subfigure}

    \vspace{1em}

    \begin{subfigure}{\textwidth}
        \centering
        \includegraphics[width=0.5\linewidth]{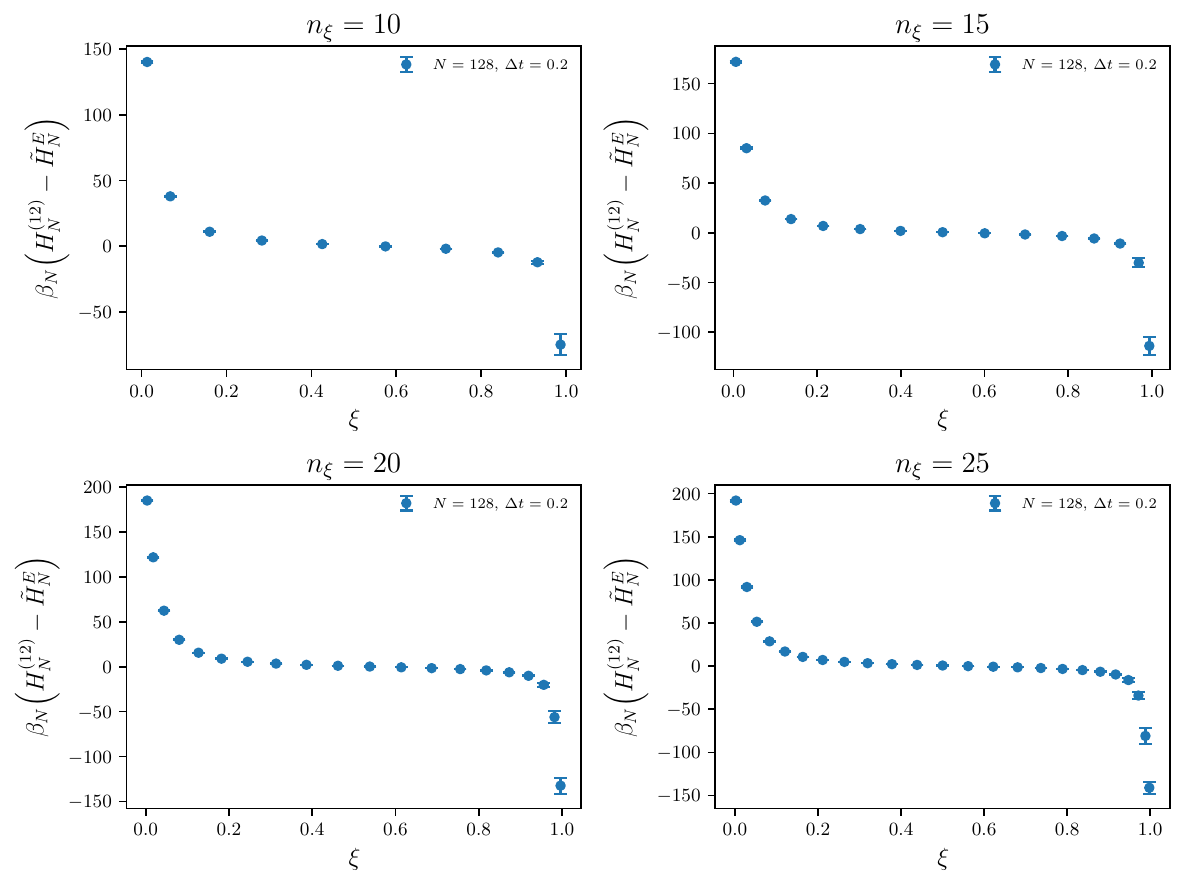}
        \caption{$\beta_N = 25$}
    \end{subfigure}

    \vspace{1em}

    \begin{subfigure}{\textwidth}
        \centering
        \includegraphics[width=0.5\linewidth]{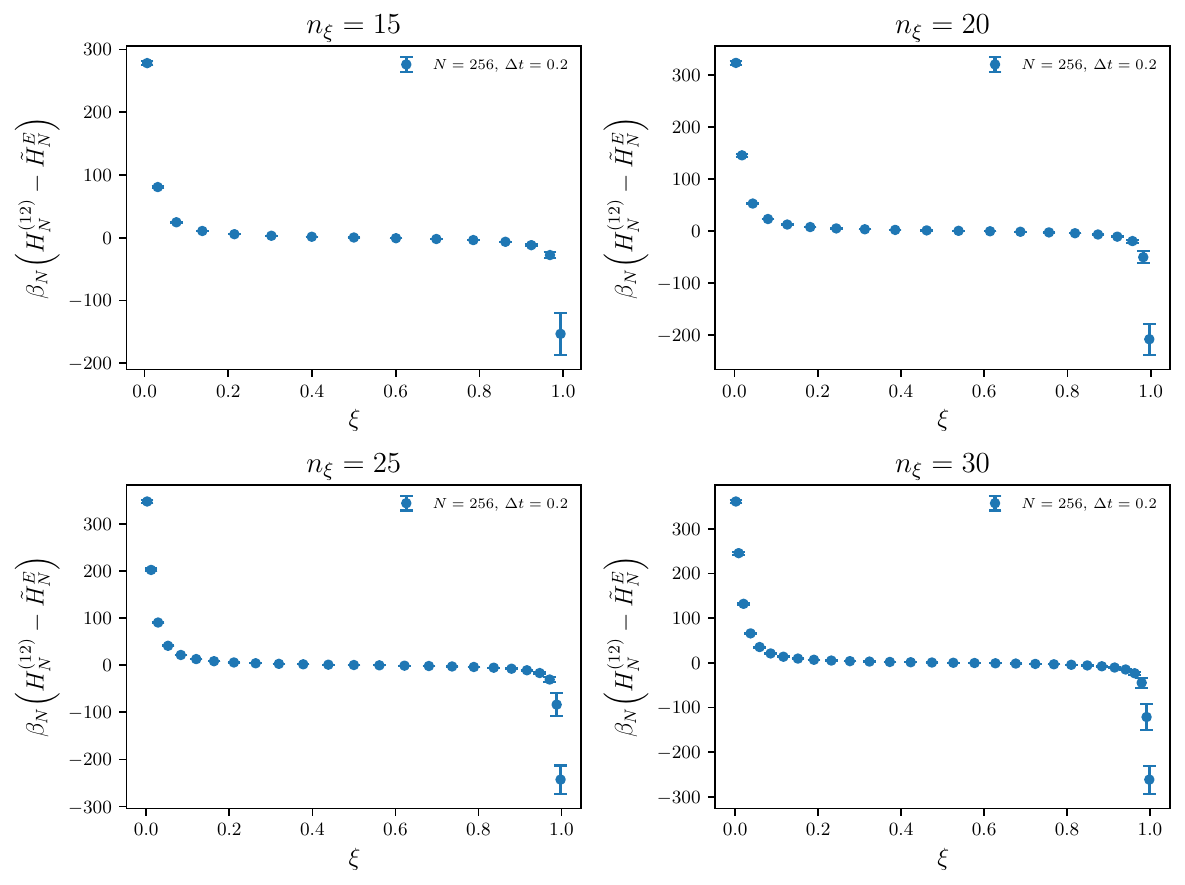}
        \caption{$\beta_N = 12$}
    \end{subfigure}
    \caption{The thermodynamic-integration plots for the $Z_{(12)}/Z_E$ simulations. Error bars have been multiplied by 500 for visibility.}
    \label{fig:TI_plots_Z12_ZE}
\end{figure}

\begin{table}[H]
\centering
\caption{ Prefactors $\langle u_P^{(J)} \rangle \times 10^{5}$ and  $\Delta F\,[\mathrm{Hartree}]$ obtained from the simulation of $Z_{(12)^*}^{(J')}/Z_{E}^{(J)}$ for various simulation parameters.
Note that the free-energy difference is very small here as the symmetry operator is similar to a rotation and does not connect two distinct versions of the molecule.
}
\begin{tabular}{ccccd{3.7}d{2.8}d{2.8}d{2.6}d{2.7}d{1.6}d{2.6}d{1.6}}
\toprule
$T\,[\mathrm{K}]$ & $N$ & $\Delta t$\,[fs] & $n_\xi$ & \head{$\Delta F\cdot 10^{5}$} & \head{$\langle u_E^{(J=0)} \rangle$} & \head{$\langle u_{(12)^*}^{(J=0)} \rangle$}& \head{$\langle u_{E}^{(J=1)}\rangle$}& \head{$\langle u_{(12)^*}^{(J=1)}\rangle$} & \head{$\langle u_{(12)^*}^{(J=2)}\rangle$} & \head{$\langle u_{(12)^*}^{(J=3)}\rangle$} & \head{$\langle u_{(12)^*}^{(J=4)}\rangle$}\\
\midrule

\multirow{11}{*}{100} 
  & \multirow{2}{*}{64}   & 0.2   & 10 & -4.(2) & 4.5737(3) & 4.5730(4) & 10.73(1) & -3.456(7) & 1.97(1) & -0.85(1) & 0.27(1)\\
  & & 0.2 & 15 & -0.2(20) & 4.5727(4) & 4.5737(4) & 10.71(1) & -3.447(7) & 1.96(1) & -0.85(1) & 0.283(9)\\
\cmidrule{2-12}
  & \multirow{4}{*}{128} & 0.2 & 10 & -3.(5) & 1.6078(2) & 1.6078(2) & 3.759(4) & -1.212(3) & 0.690(5) & -0.295(5) & 0.091(4)\\
  &  & 0.2 & 15 & -6.(4) & 1.6075(1) & 1.6078(2) & 3.773(4) & -1.211(3) & 0.687(5) & -0.295(5) & 0.096(4)\\
  &  & 0.2 & 20 & -4.(4) & 1.6080(2) & 1.6078(2) & 3.770(4) & -1.214(3) & 0.692(5) & -0.299(5) & 0.097(4) \\
  &  & 0.2 & 25 & 3.(3) & 1.6077(2) & 1.6078(2) & 3.762(5) & -1.210(3) & 0.687(5) & -0.293(5) & 0.092(3) \\
 
\cmidrule{2-12}
  & \multirow{4}{*}{256} & 0.2 & 15 & 5.(11) & 0.56695(7) & 0.56696(6) & 1.327(1) & -0.425(1) & 0.238(2) & - 0.099(2) & 0.031(1) \\
  & & 0.2& 20 & 16.(9) & 0.56698(5) & 0.56692(6) & 1.325(2) & -0.426(1) & 0.240(2) & -0.102(2) & 0.033(1)\\
  & & 0.2& 25 & -5.(9) & 0.56684(6) & 0.56698(6) & 1.328(1) & -0.4272(9) & 0.243(2) & -0.104(2) & 0.033(1)\\
  & & 0.2& 30 & 5.(8) & 0.56687(6) & 0.56693(6) & 1.326(1) & -0.425(1) & 0.239(2) & -0.101(2) & 0.033(1)\\

\bottomrule
\end{tabular}\label{tab:Ammonia_12inv_ratios_2}
\end{table}

\begin{table}[H]
\centering
\caption{ Partition-function ratios $Z_{(12)^*}^{(J')}/Z_{E}^{(J)}$ with  various values of $J'$ and $J$ obtained from different simulation parameters.}
\begin{tabular}{ccccd{2.7}d{2.6}d{1.6}d{2.6}d{1.8}d{2.7}}
\toprule
$T\,[\mathrm{K}]$ & $N$ & $\Delta t$\,[fs] & $n_\xi$ & \head{$Z_{(12)^*}^{(J=0)}/Z_{E}^{(J=0)}$} & \head{$Z_{(12)^*}^{(J=1)}/Z_{E}^{(J=0)}$} & \head{$Z_{(12)^*}^{(J=2)}/Z_{E}^{(J=0)}$}& \head{$Z_{(12)^*}^{(J=3)}/Z_{E}^{(J=0)}$}& \head{$Z_{(12)^*}^{(J=4)}/Z_{E}^{(J=0)}$} & \head{$Z_{(12)^*}^{(J=1)}/Z_{E}^{(J=1)}$}\\
\midrule

\multirow{11}{*}{100} 
  & \multirow{2}{*}{64}   & 0.2   & 10 & 1.002(1) & -0.757(2) & 0.432(3) & -0.186(3) & 0.060(2) & -0.3229(8)\\
  & & 0.2 & 15 & 1.000(1) & -0.754(2) & 0.429(3) & -0.186(2) & 0.062(2) & -0.3219(8) \\
\cmidrule{2-10}
  & \multirow{4}{*}{128} & 0.2 & 10 & 1.001(1) & -0.755(2) & 0.429(3) & -0.183(3) & 0.057(2) & -0.3228(9) \\
  &  & 0.2 & 15 & 1.002(1)& -0.754(2) & 0.428(3) & -0.184(3) & 0.060(2) & -0.321(1) \\
  &  & 0.2 & 20 & 1.001(1) & -0.755(2) & 0.431(3) & -0.186(3) & 0.060(3) & -0.3221(9)\\
  &  & 0.2 & 25 & 0.9993(8) & -0.752(2) & 0.427(3) & -0.182(3) & 0.057(2) & -0.3214(9)\\
 
\cmidrule{2-10}
  & \multirow{4}{*}{256} & 0.2 & 15 & 0.999(1) & -0.749(2) & 0.419(3) & -0.175(3) & 0.055(2) & -0.3199(9)\\
  & & 0.2& 20 & 0.998(1) & -0.750(2)& 0.423(3) & -0.180(3) & 0.057(2) & -0.3208(9)\\
  & & 0.2& 25 & 1.001(1) & -0.754(2) & 0.428(3) & -0.183(3) & 0.059(2) & -0.3220(9)\\
  & & 0.2& 30 & 0.999(1) & -0.749(2) & 0.421(3) & -0.179(3) & 0.058(2) & -0.3204(9)\\

\bottomrule
\end{tabular}\label{tab:Ammonia_12inv_ratios}
\end{table}

\begin{figure}[H]
    \centering

    \begin{subfigure}{\textwidth}
        \centering
        \includegraphics[width=0.5\linewidth]{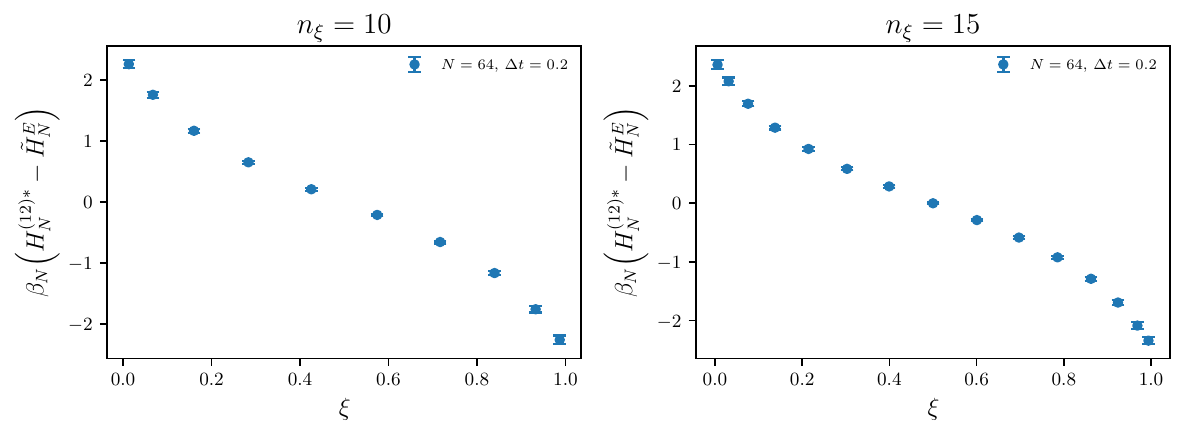}
        \caption{$\beta_N = 49$}
    \end{subfigure}

    \vspace{1em}

    \begin{subfigure}{\textwidth}
        \centering
        \includegraphics[width=0.5\linewidth]{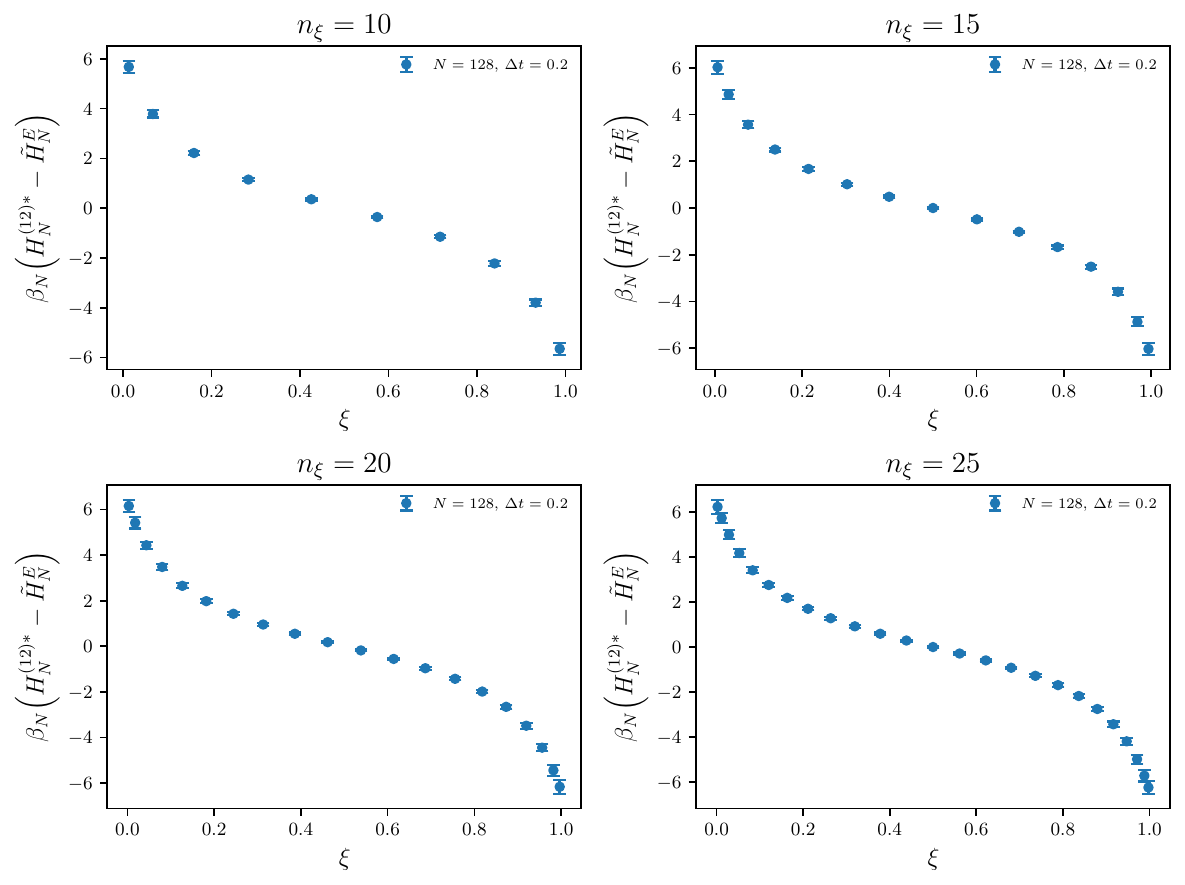}
        \caption{$\beta_N = 25$}
    \end{subfigure}

    \vspace{1em}

    \begin{subfigure}{\textwidth}
        \centering
        \includegraphics[width=0.5\linewidth]{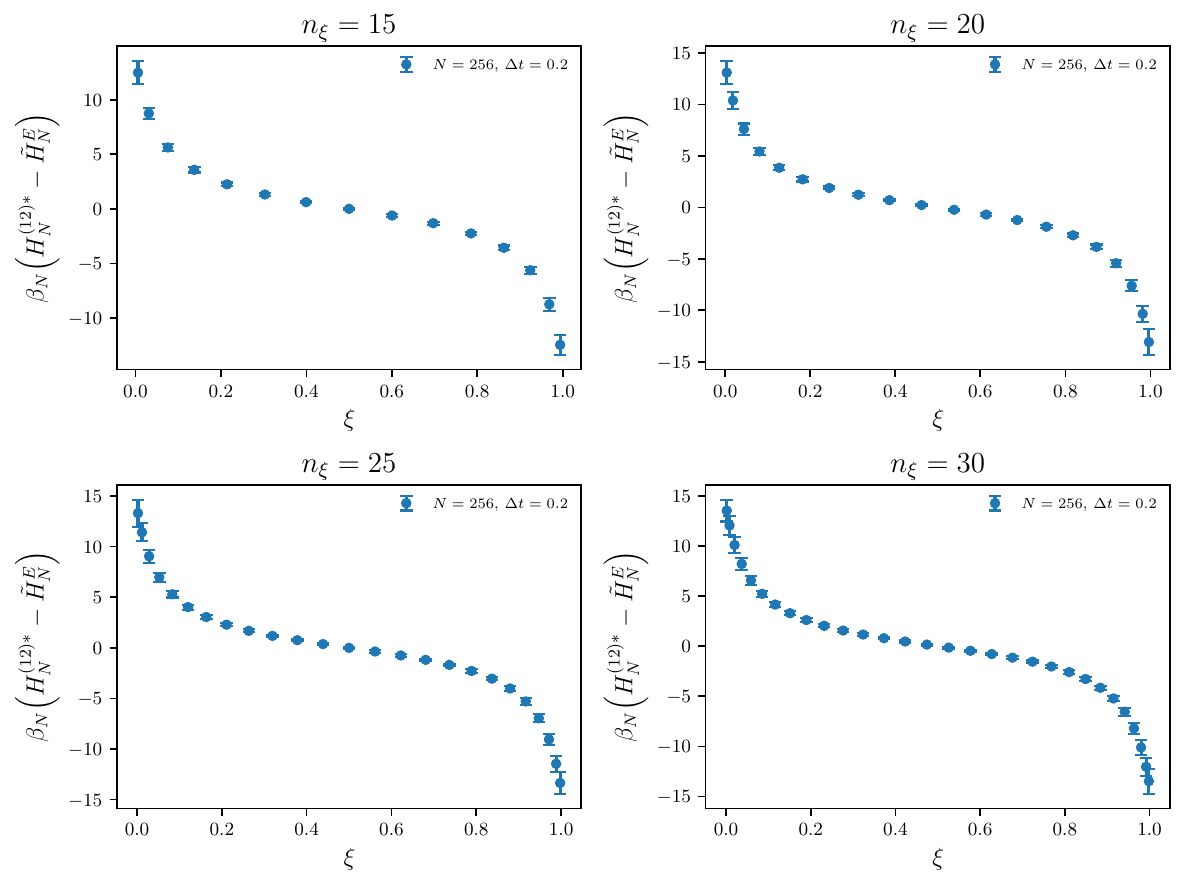}
        \caption{$\beta_N = 12$}
    \end{subfigure}
    \caption{The thermodynamic-integration plots for the $Z_{(12)^*}/Z_E$ simulations. Error bars have been multiplied by 500 for visibility.}
    \label{fig:TI_plots_Z12inv_ZE}
\end{figure}

\begin{table}[H]
\centering
\caption{ Prefactors $\langle u_P^{(J)} \rangle \times 10^{5}$ and  $\Delta F\,[\mathrm{Hartree}]$ obtained from the simulation of $Z_{(123)^*}^{(J')}/Z_{E}^{(J)}$ for various simulation parameters.}
\begin{tabular}{ccccd{2.7}d{2.8}d{2.8}d{2.6}d{2.6}d{1.6}d{2.6}d{2.6}}
\toprule
$T\,[\mathrm{K}]$ & $N$ & $\Delta t$\,[fs] & $n_\xi$ & \head{$\Delta F$} & \head{$\langle u_E^{(J=0)} \rangle $} & \head{$ \langle u_{(123)^*}^{(J=0)} \rangle$}& \head{$ \langle u_{E}^{(J=1)}\rangle$}& \head{$\langle u_{(123)^*}^{(J=1)}\rangle$} & \head{$\langle u_{(123)^*}^{(J=2)}\rangle$} & \head{$\langle u_{(123)^*}^{(J=3)}\rangle$} & \head{$\langle u_{(123)^*}^{(J=4)}\rangle$}\\
\midrule

\multirow{11}{*}{100} 
  & \multirow{2}{*}{64}   & 0.2   & 10 & 0.1045(2) & 4.5737(4) & 4.6395(7) & 10.73(1) & 7.13(1) & 1.53(2) & -2.09(2) & -1.31(2)\\
  & & 0.2 & 15 & 0.1046(2) & 4.5733(4) & 4.6383(6) & 10.72(1) & 7.13(1) & 1.50(2) & -2.12(2) & -1.31(2) \\
\cmidrule{2-12}
  & \multirow{4}{*}{128} & 0.2 & 10 & 0.2120(7) & 1.6078(2) & 1.6306(3) & 3.757(5) & 2.506(4) & 0.533(6) & -0.740(8) & -0.455(7)\\
  &  & 0.2 & 15 & 0.2107(4) & 1.6077(2) & 1.6308(3) & 3.765(4) & 2.503(4) & 0.523(7) & -0.751(8) & -0.458(7)\\
  &  & 0.2 & 20 & 0.2109(4) & 1.6078(2) & 1.6303(2) & 3.769(4) & 2.495(4) & 0.520(7) & -0.737(8) & -0.451(7)\\
  &  & 0.2 & 25 & 0.2119(4) & 1.6077(2) & 1.6302(3) & 3.771(4) & 2.504(4) & 0.527(7) & -0.747(8) & -0.452(7)\\
 
\cmidrule{2-12}
  & \multirow{4}{*}{256} & 0.2 & 15 & 0.423(1) & 0.56685(6) & 0.57481(8) & 1.327(1) & 0.881(1) & 0.182(2) & -0.268(3) & -0.165(3) \\
  & & 0.2& 20 & 0.4217(9) & 0.56685(6) & 0.57486(9) & 1.327(2) & 0.882(1) & 0.187(2) & -0.261(3) & -0.160(2) \\
  & & 0.2& 25 & 0.4241(7) & 0.56685(6) & 0.57494(9) & 1.323(2) & 0.882(1) & 0.185(2) & -0.263(3) & -0.161(2) \\
  & & 0.2& 30 & 0.4234(8) & 0.56682(6) & 0.57488(9) & 1.329(1) & 0.881(1) & 0.184(2) & -0.262(3) & -0.159(3) \\

\bottomrule
\end{tabular}\label{tab:Ammonia_123inv_ratios_2}
\end{table}

\begin{table}[H]
\centering
\caption{ Partition-function ratios $Z_{(123)^*}^{(J')}/Z_{E}^{(J)} \times 10^3$ for various values of $J'$ and $J$ obtained from different simulation parameters.}
\begin{tabular}{ccccd{3.5}d{2.5}d{1.5}d{2.5}d{2.5}d{1.5}}
\toprule
$T\,[\mathrm{K}]$ & $N$ & $\Delta t$\,[fs] & $n_\xi$ & \head{$Z_{(123)^*}^{(J=0)}/Z_{E}^{(J=0)}$} & \head{$Z_{(123)^*}^{(J=1)}/Z_{E}^{(J=0)}$} & \head{$Z_{(123)^*}^{(J=2)}/Z_{E}^{(J=0)}$}& \head{$Z_{(123)^*}^{(J=3)}/Z_{E}^{(J=0)}$}& \head{$Z_{(123)^*}^{(J=4)}/Z_{E}^{(J=0)}$} & \head{$Z_{(123)^*}^{(J=1)}/Z_{E}^{(J=1)}$}\\
\midrule

\multirow{11}{*}{100} 
  & \multirow{2}{*}{64}   & 0.2   & 10 & 5.85(6) & 9.0(1) & 1.92(3) & -2.64(4) & -1.65(3) & 3.83(4)\\
  & & 0.2 & 15 & 5.81(5) & 8.92(9) & 1.88(3) & -2.66(4) & -1.64(3) & 3.81(4)\\
\cmidrule{2-10}
  & \multirow{4}{*}{128} & 0.2 & 10 & 5.42(9) & 8.3(2) & 1.77(3) & -2.46(5) & -1.51(3) & 3.57(6) \\
  &  & 0.2 & 15 & 5.61(6) & 8.60(9) & 1.80(3) & -2.58(4) & -1.57(3) & 3.68(4) \\
  &  & 0.2 & 20 & 5.58(5) & 8.54(8) & 1.78(3) & -2.52(4) & -1.54(3) & 3.64(4)\\
  &  & 0.2 & 25 & 5.43(5) & 8.35(7) & 1.76(3) & -2.49(3) & -1.51(3) & 3.56(3) \\
 
\cmidrule{2-10}
  & \multirow{4}{*}{256} & 0.2 & 15 & 5.51(8) & 8.5(1) & 1.74(3) & -2.57(5) & -1.58(3) & 3.61(5)\\
  & & 0.2& 20 & 5.58(6) & 8.57(9) & 1.82(3) & -2.53(4) & -1.55(3) & 3.66(4) \\
  & & 0.2& 25 & 5.43(5) & 8.32(7) & 1.74(3) & -2.49(3) & -1.52(2) & 3.57(3) \\
  & & 0.2& 30 & 5.47(6) & 8.38(8) & 1.75(3) & -2.50(4) & -1.51(3) & 3.57(4)\\

\bottomrule
\end{tabular}\label{tab:Ammonia_123inv_ratios}
\end{table}

\begin{figure}[H]
    \centering

    \begin{subfigure}{\textwidth}
        \centering
        \includegraphics[width=0.5\linewidth]{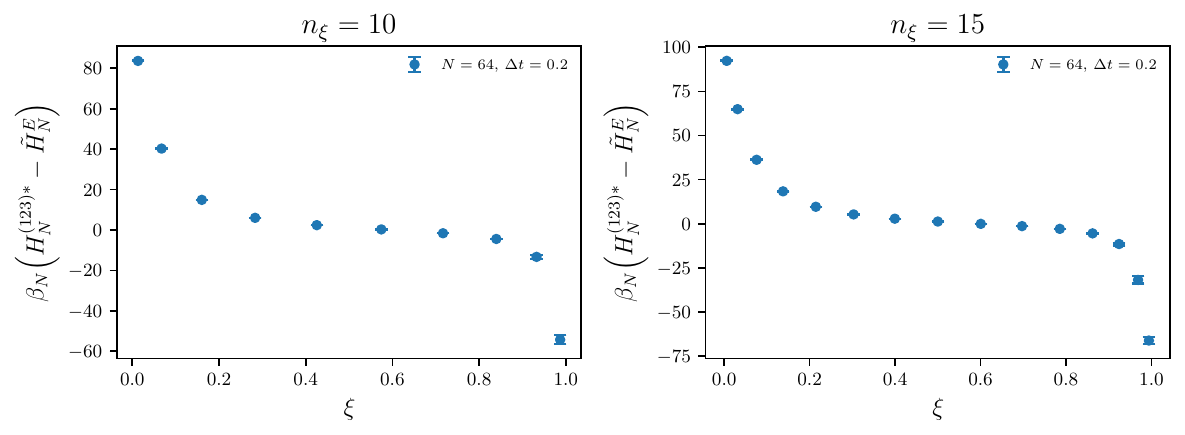}
        \caption{$\beta_N = 49$}
    \end{subfigure}

    \vspace{1em}

    \begin{subfigure}{\textwidth}
        \centering
        \includegraphics[width=0.5\linewidth]{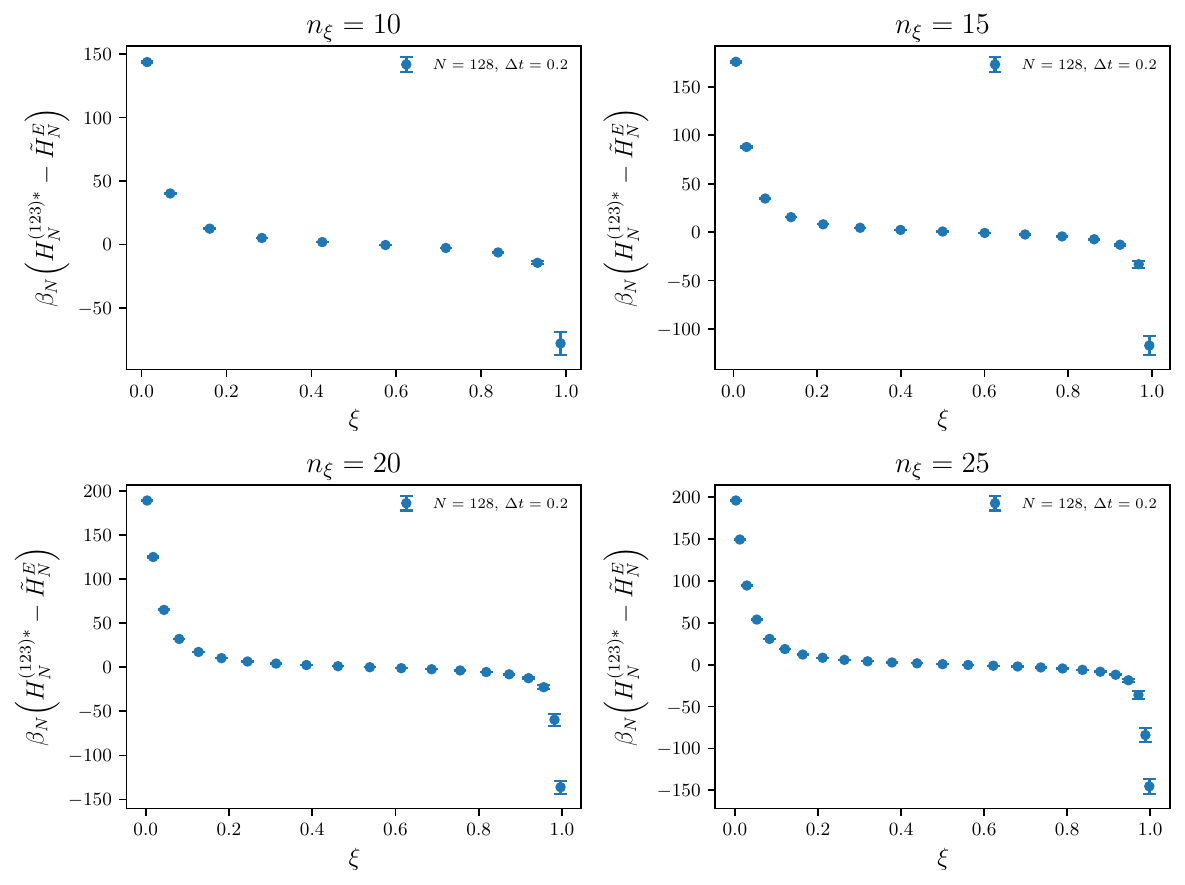}
        \caption{$\beta_N = 25$}
    \end{subfigure}

    \vspace{1em}
    
    \begin{subfigure}{\textwidth}
        \centering
        \includegraphics[width=0.5\linewidth]{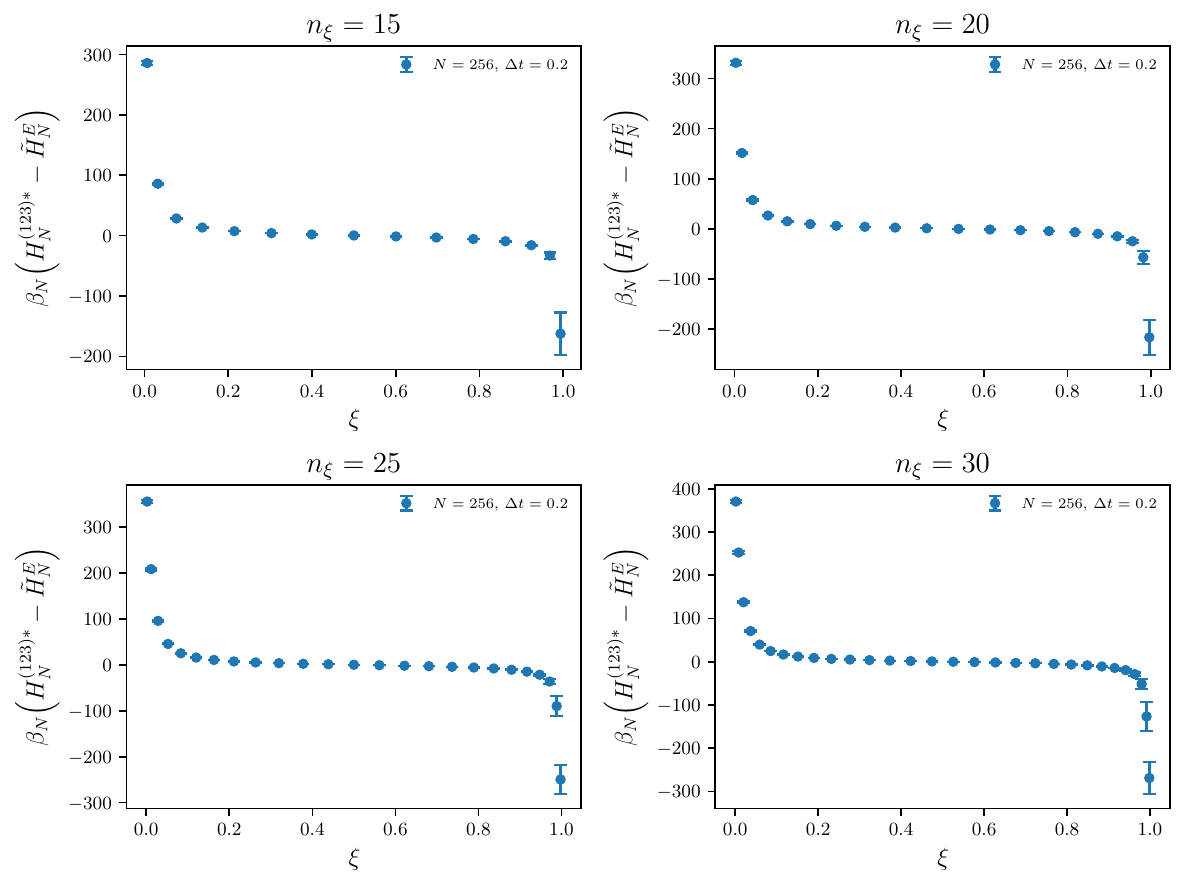}
        \caption{$\beta_N = 12$}
    \end{subfigure}
    \caption{The thermodynamic-integration plots for the $Z_{(123)^*}/Z_E$ simulations. Error bars have been multiplied by 500 for visibility.}
    \label{fig:TI_plots_Z123inv_ZE}
\end{figure}

\begin{sidewaystable}[htbp]
\centering
\caption{ Prefactors $\langle u_P^{(J)} \rangle \times 10^{5}$ and  $\Delta F\,[\mathrm{Hartree}]$ obtained from the simulation of $Z_{(12)}^{(J)}/Z_{(12)^*}^{(J)}$ for various simulation parameters.}
\begin{tabular}{ccccd{2.7}d{1.8}d{2.8}d{1.6}d{2.6}d{1.6}d{1.8}d{2.8}d{1.6}d{2.6}d{1.6}}
\toprule
$T\,[\mathrm{K}]$ & $N$ & $\Delta t$\,[fs] & $n_\xi$ & \head{$\Delta F$} & \head{$\langle u^{(J=0)}_{(12)^*}\rangle$} & \head{$\langle u^{(J=1)}_{(12)^*}\rangle$}& \head{$\langle u^{(J=2)}_{(12)^*}\rangle$}& \head{$\langle u^{(J=3)}_{(12)^*}\rangle$} & \head{$\langle u^{(J=4)}_{(12)^*}\rangle$} & \head{$\langle u^{(J=0)}_{(12)}\rangle$} & \head{$\langle u^{(J=1)}_{(12)}\rangle$}& \head{$\langle u^{(J=2)}_{(12)}\rangle$}& \head{$\langle u^{(J=3)}_{(12)}\rangle$} & \head{$\langle u^{(J=4)}_{(12)}\rangle$}\\
\midrule

\multirow{11}{*}{100} 
  & \multirow{2}{*}{64}   & 0.2   & 10 & 0.1046(2) & 4.5733(4) & -3.453(8)& 1.97(1)& -0.84(1) & 0.27(1) & 4.6389(7) & -3.452(8) & 1.91(1) & -0.78(1) & 0.24(1)\\
  &  & 0.2 & 15 & 0.1047(2) & 4.5748(4) & -3.456(7) & 1.97(1) & -0.85(1) & 0.27(1) & 4.6405(7) & -3.458(8) & 1.93(1) & -0.81(1) & 0.26(1) \\
\cmidrule{2-15}
 & \multirow{4}{*}{128} &0.2  & 10 & 0.2119(5) & 1.6076(2) & -1.206(3) & 0.679(4)& -0.288(4)& 0.092(3) & 1.6306(2) & -1.213(3) & 0.673(5) & -0.279(5) & 0.088(3)\\
 &  &  0.2  & 15 & 0.2118(4) & 1.6079(2) & -1.208(2) & 0.681(4) & -0.287(4) & 0.088(3) & 1.6302(2) & -1.213(3) & 0.672(5) & -0.278(4) & 0.087(3) \\
& & 0.2 & 20 & 0.2117(4) & 1.6081(2) & -1.207(2) & 0.681(4) & -0.289(4) & 0.091(3) & 1.6308(2) & -1.211(3) & 0.668(5) & -0.275(5) & 0.085(3) \\
& & 0.2 & 25 & 0.2118(3) & 1.6077(2) & -1.211(3) & 0.687(5) & -0.292(5) & 0.093(4)& 1.6306(2) & -1.216(2) & 0.677(4) & -0.282(4) & 0.087(4)\\
 
\cmidrule{2-15}
  & \multirow{4}{*}{256} & 0.2 & 15 & 0.422(1) & 0.56693(6) & -0.4286(7) & 0.245(1)& -0.107(1) & 0.037(1)& 0.57478(9) & -0.427(1) & 0.236(2) & -0.097(2) & 0.030(1) \\
  & & 0.2& 20 & 0.4243(9) & 0.56676(6) & -0.4270(10) & 0.243(2)& -0.104(2) & 0.033(1)& 0.57496(9) & -0.4272(10) & 0.236(2) & -0.097(2) & 0.030(1) \\
  & & 0.2& 25 & 0.4236(8) & 0.56690(9) & -0.4261(9)& 0.241(2)& -0.103(2)& 0.034(1) & 0.57470(9) & -0.428(1) & 0.237(2) & -0.099(2) & 0.031(1)\\
  & & 0.2& 30 & 0.4248(6) & 0.56682(6) & -0.4262(9) & 0.241(1) & -0.102(1) & 0.032(1)& 0.5749(1) & -0.4283(1) & 0.237(2) & -0.097(2) & 0.029(1)\\
\bottomrule
\end{tabular}\label{tab:Ammonia_res_new_supl2}
\end{sidewaystable}

\begin{table}[H]
\centering
\caption{ Partition-function ratios $Z_{(12)}^{(J)}/Z_{(12)^*}^{(J)} \times 10^3$ for various values of $J$ obtained from different simulation parameters.}
\begin{tabular}{ccccd{2.5}d{1.5}d{1.5}d{1.4}d{1.4}}
\toprule
$T\,[\mathrm{K}]$ & $N$ & $\Delta t$\,[fs] & $n_\xi$ & \head{$J=0$} & \head{$J=1$} & \head{$J=2$}& \head{$J=3$}& \head{$J=4$}\\
\midrule

\multirow{11}{*}{100} 
  & \multirow{2}{*}{64}   & 0.2   & 10 & 5.82(6)& 5.73(7) & 5.57(8) & 5.3(1) & 5.0(3)\\
  &  & 0.2 & 15 & 5.80(6) & 5.72(6) & 5.59(8) & 5.4(1) & 5.4(3)\\
\cmidrule{2-9}
  & \multirow{4}{*}{128} &0.2  & 10 & 5.44(7) & 5.40(8) & 5.32(9) & 5.2(1) & 5.2(3)\\
 &  &  0.2  & 15 & 5.45(5) & 5.40(5) & 5.31(7) & 5.2(1) & 5.3(3)\\
& & 0.2 & 20 & 5.47(5) & 5.41(5) & 5.30(7)& 5.1(1)& 5.0(3)\\
& & 0.2 & 25 & 5.46(5)& 5.41(5)& 5.31(7)& 5.2(1)& 5.1(3)\\
 
\cmidrule{2-9}
  & \multirow{4}{*}{256} & 0.2 & 15 & 5.56(8) & 5.47(8)& 5.29(9)& 5.0(1)& 4.4(3)\\
  & & 0.2& 20 & 5.41(6) & 5.33(6) & 5.19(8) & 5.0(1) & 4.9(3) \\
  & & 0.2& 25 & 5.45(6) & 5.39(6) & 5.29(7) & 5.1(1) & 4.9(3)\\
  & & 0.2& 30 & 5.38(4) & 5.33(4) & 5.11(6) & 5.0(1) & 4.7(3) \\

\hline
\end{tabular}\label{tab:Ammonia_res_new_supl1}
\end{table}

\begin{figure}[H]
    \centering
    \begin{subfigure}{\textwidth}
        \centering
        \includegraphics[width=0.5\linewidth]{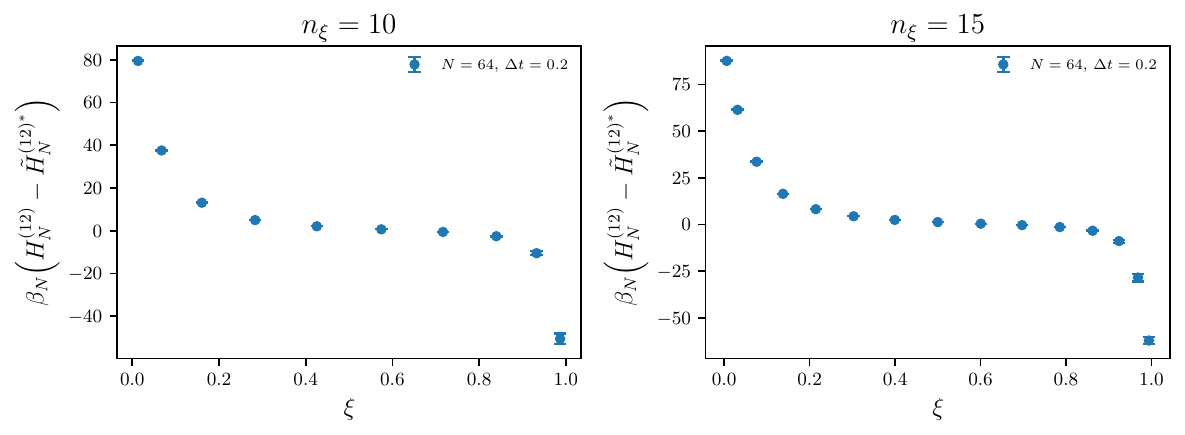}
        \caption{$\beta_N = 49$}
    \end{subfigure}

    \vspace{1em}

    \begin{subfigure}{\textwidth}
        \centering
        \includegraphics[width=0.5\linewidth]{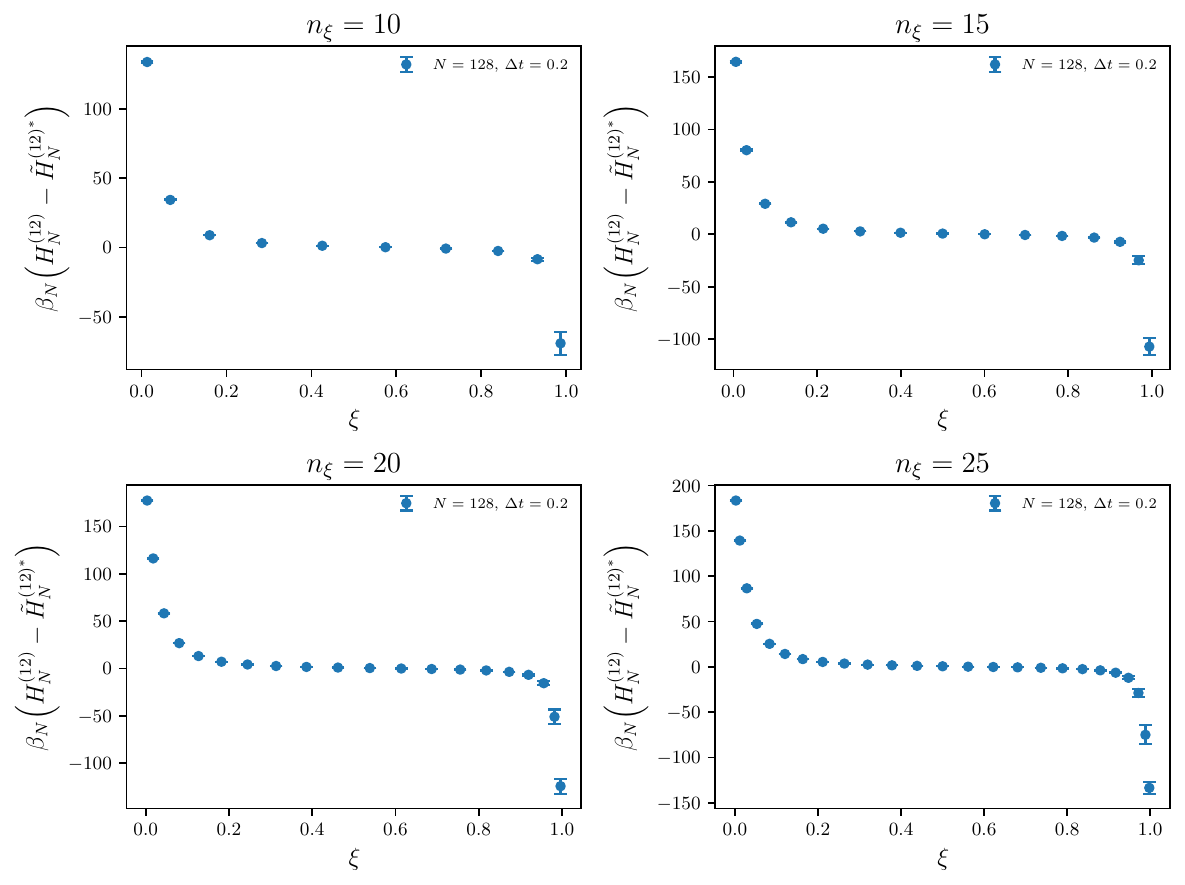}
        \caption{$\beta_N = 25$}
    \end{subfigure}

    \vspace{1em}

    \begin{subfigure}{\textwidth}
        \centering
        \includegraphics[width=0.5\linewidth]{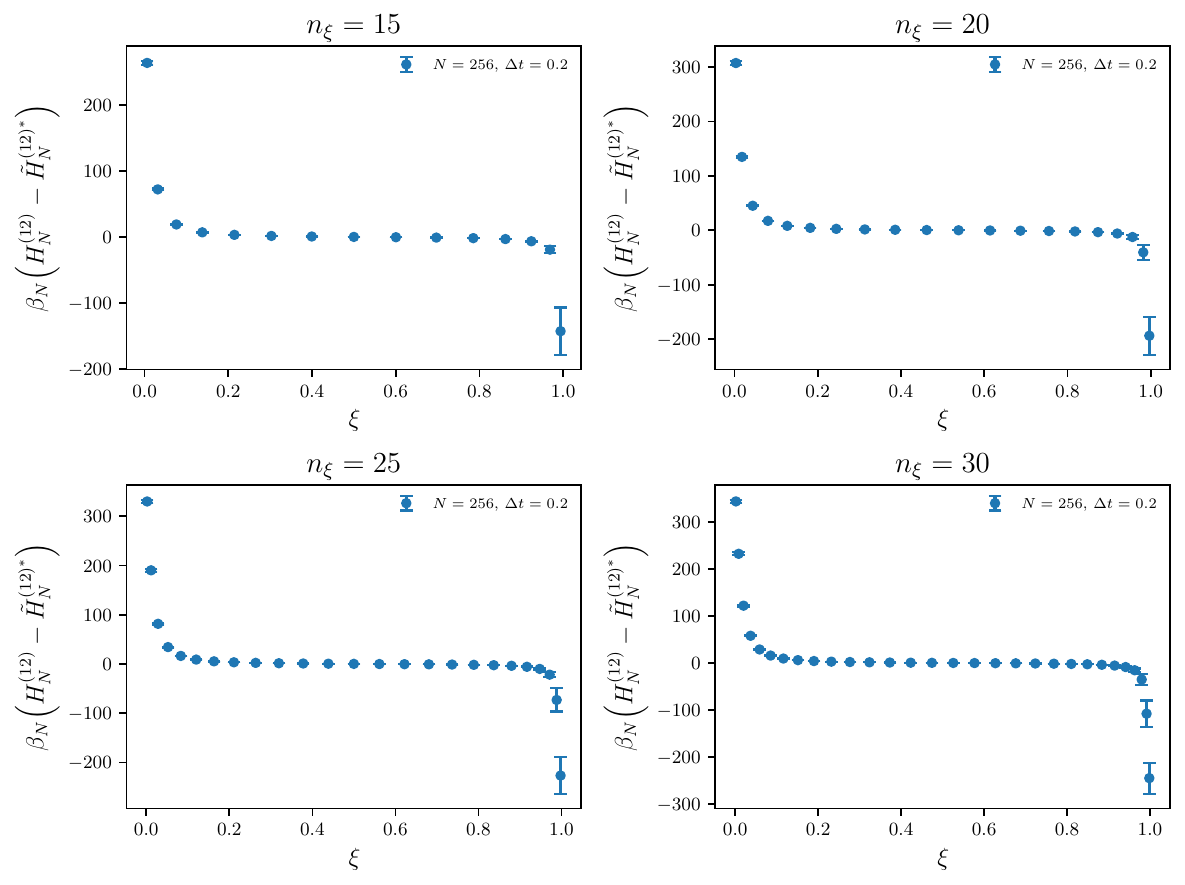}
        \caption{$\beta_N = 12$}
    \end{subfigure}
    \caption{The thermodynamic-integration plots for the $Z_{(12)}/Z_{(12)^*}$ simulations. Error bars have been multiplied by 500 for visibility.}
    \label{fig:TI_plots_Z12_Z12*}
\end{figure}

\newpage
\section{Error propagation for tunneling splittings}
Consider a general multivariate function $f(\mathbf{x})$, where $\mathbf{x}=(x_1,x_2,\cdots)$ is a set of random variables with means $\bar{\mathbf{x}}=(\bar{x}_1,\bar{x}_2,\cdots)$.
In the limit where the fluctuations of $\mathbf{x}$ are small, the fluctuation of $f(\mathbf{x})$ can be approximated by the first-order Taylor expansion about $\bar{\mathbf{x}}$,
\begin{equation}
\Delta f \equiv f(\bar{\mathbf{x}}+\Delta\mathbf{x})-f(\bar{\mathbf{x}})\approx 
\sum_{k}\frac{\partial f}{\partial x_k}\Delta x_k,
\end{equation}
where all partial derivatives are evaluated at $\mathbf{x}=\bar{\mathbf{x}}$.
Accordingly, the standard deviation of $f(\mathbf{x})$ is given by
\begin{equation}\label{eq:error-cov}
\sigma_f = \sqrt{\sum_{kk'}\frac{\partial f}{\partial x_k}\frac{\partial f}{\partial x_{k'}}C_{kk'}},
\end{equation}
where $C_{kk'}$ is the covariance between $x_k$ and $x_{k'}$.
In the special case where the random variables are statistically independent, Eq.~\eqref{eq:error-cov} reduces to
\begin{equation}\label{eq:error-sigma}
\sigma_f = \sqrt{\sum_{k}\left(\frac{\partial f}{\partial x_k}\sigma_k\right)^2},
\end{equation}
where $\sigma_k=\sqrt{C_{kk}}$ is the standard deviation of $x_k$.

In our tunneling-splitting calculations, Eq.~\eqref{eq:error-sigma} is used to estimate the stochastic uncertainty of the splitting.
Specifically, the function $f$ is the tunneling-splitting expression obtained from Eq.~(13) by dividing both the numerator and the denominator by a common reference partition function, which allows the splitting to be written solely in terms of partition-function ratios.
The variables $\{x_k\}$ correspond to these ratios.
For a given tunneling splitting, the statistical uncertainties of all the relevant ratios are evaluated using the formulas presented in the Appendix of Ref.~\onlinecite{PIMDtunnel}.
These quantities provide the $\sigma_k$ or $C_{kk'}$ entering Eqs.~\eqref{eq:error-cov} and \eqref{eq:error-sigma}.
The derivatives required for the error propagation are obtained directly from the rewritten form of Eq.~(13) and evaluated at the mean values of the ratios.
This procedure yields the final stochastic uncertainty of the tunneling splitting.

Note that using Eq.~\eqref{eq:error-sigma} for the error propagation is rigorous in the ammonia case, as all the ratios entering the tunneling-splitting expressions are from different sets of simulations and are therefore statistically independent.
However, in the water case, the relevant ratios are not all statistically independent as some of them are calculated from the same set of trajectories.
For simplicity, we still applied Eq.~\eqref{eq:error-sigma} to water.
As such, the statistical errors reported in Table~II and Table~III in the main text could be underestimated or overestimated by up to a factor of $\sqrt{2}$, which is small enough that it would not change our conclusions. %

\section*{References}
\bibliography{new_refs,references}